\documentclass[11pt,a4paper]{article}
\usepackage{jcappub}
\usepackage{lmodern}
\usepackage{graphicx,color}
\graphicspath{{figures/}}
\usepackage{comment}
\usepackage{cancel}

\usepackage[T1]{fontenc}
\usepackage[utf8]{inputenc}
\usepackage{color}
\usepackage{amsmath}
\usepackage{amssymb}
\usepackage{esint}
\usepackage{url}
\usepackage[normalem]{ulem}

\makeatletter

\definecolor{somegreen}{cmyk}{0,0.49,0.98,0.09}
\definecolor{red}{rgb}{1,0,0}
\definecolor{magenta}{cmyk}{0,1,0,0}
\definecolor{lavender}{cmyk}{0.16,0.67,0,0.57}
\definecolor{darkgreen}{rgb}{0,0.65,0.05}
\definecolor{antiquefuchsia}{rgb}{0.33, 0.1, 0.89}

\def\beqra{\begin{eqnarray}}
\def\eeqra{\end{eqnarray}}
\def\beq{\begin{equation}}
\def\eeq{\end{equation}}

\def\ds{\displaystyle}

\def\bx{{\bf{x}}}

\def\bV0{{\bf{V_0}}}
\def\re#1{(\ref{#1})}

\def\bx{{\bf{x}}}

\def\bs{{\bf{s}}}

\def\bV{{\bf{V}}}

\def\cov{\mathbf{C}}
\def\btheta{\boldsymbol{\theta}}
\def\bs{\mathbf{s}}

\newcommand{\diff}{\mathop{}\!\mathrm{d}}
\newcommand{\hMpc}{h\,\mathrm{Mpc}^{-1}}

\newcommand{\kmax}{k_\mathrm{max}}
\newcommand{\Ngrid}{N_\mathrm{grid}}

\newcommand{\fNL}{f_\mathrm{NL}}
\newcommand{\fNLloc}{f_\mathrm{NL}^\mathrm{local}}
\newcommand{\fNLeq}{f_\mathrm{NL}^\mathrm{equil}}
\newcommand{\fNLort}{f_\mathrm{NL}^\mathrm{ortho}}
\newcommand{\Mmin}{M_\mathrm{min}}

\newcommand{\Planck}{\textit{Planck}}
\newcommand{\Quijote}{\textsc{Quijote}}
\newcommand{\QuijotePNG}{\textsc{Quijote-png}}

\let\jnl@style=\rm
\def\ref@jnl#1{{\jnl@style#1}}

\def\aj{\ref@jnl{AJ}}                   
\def\actaa{\ref@jnl{Acta Astron.}}      
\def\araa{\ref@jnl{ARA\&A}}             
\def\apj{\ref@jnl{ApJ}}                 
\def\apjl{\ref@jnl{ApJ}}                
\def\apjs{\ref@jnl{ApJS}}               
\def\ao{\ref@jnl{Appl.~Opt.}}           
\def\apss{\ref@jnl{Ap\&SS}}             
\def\aap{\ref@jnl{A\&A}}                
\def\aapr{\ref@jnl{A\&A~Rev.}}          
\def\aaps{\ref@jnl{A\&AS}}              
\def\azh{\ref@jnl{AZh}}                 
\def\baas{\ref@jnl{BAAS}}               
\def\bac{\ref@jnl{Bull. astr. Inst. Czechosl.}}
\def\caa{\ref@jnl{Chinese Astron. Astrophys.}}
\def\cjaa{\ref@jnl{Chinese J. Astron. Astrophys.}}
\def\icarus{\ref@jnl{Icarus}}           
\def\jcap{\ref@jnl{J. Cosmology Astropart. Phys.}}
\def\jrasc{\ref@jnl{JRASC}}             
\def\memras{\ref@jnl{MmRAS}}            
\def\mnras{\ref@jnl{MNRAS}}             
\def\na{\ref@jnl{New A}}                
\def\nar{\ref@jnl{New A Rev.}}          
\def\pra{\ref@jnl{Phys.~Rev.~A}}        
\def\prb{\ref@jnl{Phys.~Rev.~B}}        
\def\prc{\ref@jnl{Phys.~Rev.~C}}        
\def\prd{\ref@jnl{Phys.~Rev.~D}}        
\def\pre{\ref@jnl{Phys.~Rev.~E}}        
\def\prl{\ref@jnl{Phys.~Rev.~Lett.}}    
\def\pasa{\ref@jnl{PASA}}               
\def\pasp{\ref@jnl{PASP}}               
\def\pasj{\ref@jnl{PASJ}}               
\def\rmxaa{\ref@jnl{Rev. Mexicana Astron. Astrofis.}}%
\def\qjras{\ref@jnl{QJRAS}}             
\def\skytel{\ref@jnl{S\&T}}             
\def\solphys{\ref@jnl{Sol.~Phys.}}      
\def\sovast{\ref@jnl{Soviet~Ast.}}      
\def\ssr{\ref@jnl{Space~Sci.~Rev.}}     
\def\zap{\ref@jnl{ZAp}}                 
\def\nat{\ref@jnl{Nature}}              
\def\iaucirc{\ref@jnl{IAU~Circ.}}       
\def\aplett{\ref@jnl{Astrophys.~Lett.}} 
\def\apspr{\ref@jnl{Astrophys.~Space~Phys.~Res.}}
\def\bain{\ref@jnl{Bull.~Astron.~Inst.~Netherlands}}
\def\fcp{\ref@jnl{Fund.~Cosmic~Phys.}}  
\def\gca{\ref@jnl{Geochim.~Cosmochim.~Acta}}   
\def\grl{\ref@jnl{Geophys.~Res.~Lett.}} 
\def\jcp{\ref@jnl{J.~Chem.~Phys.}}      
\def\jgr{\ref@jnl{J.~Geophys.~Res.}}    
\def\jqsrt{\ref@jnl{J.~Quant.~Spec.~Radiat.~Transf.}}
\def\memsai{\ref@jnl{Mem.~Soc.~Astron.~Italiana}}
\def\nphysa{\ref@jnl{Nucl.~Phys.~A}}   
\def\physrep{\ref@jnl{Phys.~Rep.}}   
\def\physscr{\ref@jnl{Phys.~Scr}}   
\def\planss{\ref@jnl{Planet.~Space~Sci.}}   
\def\procspie{\ref@jnl{Proc.~SPIE}}   

\makeatother

\begin{document}
\title{Constraining Primordial Non-Gaussianity from Large Scale Structure with the Wavelet Scattering Transform
}
\author[a]{Matteo Peron,}
\author[b]{Gabriel Jung,}
\author[c]{Michele Liguori,}
\author[a]{Massimo Pietroni}
\affiliation[a]{Dipartimento di Scienze Matematiche, Fisiche e Informatiche, Universit\`a di Parma, and INFN Gruppo Collegato di Parma, Parco Area delle Scienze 7/A, I-43124, Parma, Italy}
\affiliation[b]{Universit\'e Paris-Saclay, CNRS, Institut d’Astrophysique Spatiale, F-91405, Orsay, France}
\affiliation[c]{Dipartimento di Fisica e Astronomia G. Galilei, Universit\`a degli Studi di Padova and INFN, Sezione di Padova, Via Marzolo 8, I-35131, Padova, Italy}

\abstract{We investigate the Wavelet Scattering Transform (WST) as a tool for the study of Primordial non-Gaussianity (PNG) in Large Scale Structure (LSS), and compare its performance with that achievable via
a joint analysis with power spectrum and bispectrum (P+B). We consider the three main primordial bispectrum shapes – local, equilateral and orthogonal – and produce Fisher forecast for the corresponding $f_{\rm NL}$ amplitude
parameters, jointly with standard cosmological parameters. We analyze simulations from the publicly available $\Quijote$ and $\QuijotePNG$ N-body suites, studying both  the dark matter and halo fields.
We find that the WST outperforms the power spectrum alone on all parameters, both on the $f_{\rm NL}$'s and on cosmological ones. In particular, on $\fNLloc$ for halos, the improvement is about 27\%. 
When B is combined with P, halo constraints  from WST are weaker for $\fNLloc$ (at $\sim \, 15\%$ level), but stronger for $\fNLeq$ ($\sim \, 25\%$)  and $\fNLort$ ($\sim \, 28\%$).
Our results show that WST, both alone and in combination with P+B, can improve the extraction of information on PNG from LSS data over the one attainable by a standard P+B analysis. Moreover, we identify a  class of WST 
in which the  origin of the extra information on PNG can be cleanly isolated.}
\emailAdd{matteo.peron@unipr.it} 
\emailAdd{gabriel.jung@universite-paris-saclay.fr} 
\emailAdd{michele.liguori@unipd.it} 
\emailAdd{massimo.pietroni@unipr.it}

\maketitle


\section{Introduction}
An important goal of observational cosmology in the coming years will be the development of methodologies to extract maximum information from non-linear scales in galaxy surveys. Clearly, this will involve going beyond power spectrum analysis, since the power spectrum alone cannot capture phase information and is therefore not sensitive to structures such as, for example, filaments in the cosmic web. A natural choice in this respect is to consider the bispectrum (three-point function of the Fourier modes of the density field), as this is the lowest order non-Gaussian (NG) correlator. Indeed, many joint power spectrum and bispectrum (P$+$B) Large Scale Structure (LSS) forecasts and analyses are already present in the cosmological literature \cite{DAmico:2022gki,DAmico:2022osl,Philcox:2022frc, Ivanov:2023qzb,Hahn:2023kky} and this kind of approach to the analysis of future galaxy survey data presents some advantages, notably the availability of several well established and validated P$+$B analysis pipelines and the theoretical interpretability of bispectrum measurements -- at mildly non-linear scales -- in cosmological perturbation theory. On the other hand, the bispectrum clearly does not exhaust the available information at fully non-linear scales and a straightforward extension to higher order correlators becomes increasingly more cumbersome and difficult to implement, especially when considering the need for accurate evaluation of covariance matrices in the non-linear regime of interest and for inclusion of realistic observational effects (e.g., window functions). This motivates the search for different NG summary statistics. The ideal properties that we would require from them would be ease of implementation, efficiency at extracting all the available NG information using a limited number of modes and clear physical interpretability. With such goals in mind, several NG summaries have already been considered in LSS analysis. Among those, let us mention for example marked spectra \cite{Sheth:2005ai,Sheth:2005aj,Philcox:2020fqx,Massara:2020pli,Massara:2022zrf}, power spectra from cosmic web environments \cite{Bonnaire:2021sie, Bonnaire:2022ocm}, skew spectra \cite{Munshi:2009ik,MoradinezhadDizgah:2019xun,Schmittfull:2020hoi,Hou:2024blc}, Minkowski functionals \cite{2012MNRAS.423.3209P,2022JCAP...07..045L}, k-nearest neighbour cumulative distributions functions \cite{Banerjee:2020umh, Coulton:2023ouk}, probability distribution function (PDF) of late-time cosmic density fluctuations \cite{Friedrich:2019byw, Uhlemann:2019gni}, halo mass function probes \cite{1986ApJ...310L..21M,Matarrese:2000iz,Robinson:1999wh, LoVerde:2011iz, Bayer:2021iyb, Jung:2023kjh}, and the void abundance \cite{2009JCAP...01..010K, 2011PhRvD..83b3521D}. 

Here, we focus on another statistic that has received significant recent attention in observational cosmology, namely the Wavelet Scattering Transform (WST) \cite{https://doi.org/10.1002/cpa.21413, kymatio}. The WST has the structure of a three-layer convolutional network, in which the operations performed at each stage are, in sequence, non-linear activation via the modulus function, convolution by a wavelet filter and low-pass spatial averaging of the data. The WST has recently been applied to simulations  in \citep{Valogiannis:2021chp, Eickenberg:2022qvy,  Valogiannis:2022xwu, Blancard:2023iab,Valogiannis:2023mxf}, showing that it could efficiently capture NG information and produce significant improvements over power-spectrum-only cosmological constraints, especially on neutrino mass. 
The main goal of this work is to investigate the WST as a tool for the study of primordial non-Gaussianity (PNG) and to compare in detail its performance with that achievable via a P$+$B analysis, along with providing a general physical interpretation for this comparison. 

Non-Gaussian features in the primordial cosmological density field are a distinctive  prediction of inflationary models beyond standard single-field slow-roll. Primordial non-Gaussian signatures are encoded in the primordial bispectrum of cosmological perturbations: different deviations from the single-field slow-roll scenario produce specific functional dependencies of the bispectrum on triangular configurations of Fourier modes. This model dependence makes PNG a very powerful tool to constrain inflation. At the same time, the expected PNG signal is very small and strongly degenerate, at non-linear scales, with the bispectrum produced at late times by gravitational evolution of cosmic structures and galaxy bias. Beyond-bispectrum statistics, such as the WST, can help breaking this degeneracy, which is one of the main motivations of the present analysis.
Let us note here that, instead of relying on summary statistics, an alternative, more ambitious approach consists in performing forward modeling field level inference. Field level analysis is more and more employed in cosmological data analysis and carries the promise of extracting all the available information in the data (see, e.g., \citep{2023MNRAS.520.5746A,Stadler:2023hea,Alsing:2019xrx,Ribli:2019wtw,Ntampaka:2019ole, Jeffrey:2020itg,Villanueva-Domingo:2022rvn,Makinen:2022jsc,Shao:2022mzk, deSanti:2023zzn,Roncoli:2023ysi,Hahn:2022zxa,Lemos:2023myd,Nguyen:2024yth}). The more traditional approach, based on preliminary compression of the data in a set of summaries, albeit lossy, remains nonetheless a valid alternative, often leading to easier physical interpretation of the results and comparison with analytical models; it is also relevant as it gives insight on where (i.e., in which domain, at which scales, and so on) the bulk of information about the relevant signals lies and on how it can be efficiently recovered. 

In our work, we consider the three main primordial bispectrum shapes -- local, equilateral and orthogonal -- and produce Fisher forecast for the corresponding $f_{\rm NL}$ amplitude parameters, jointly with standard cosmological parameters ($\sigma_8$, $\Omega_m$, $n_s$, $h$). To build the Fisher matrix up to a scale $\kmax = 0.5 \, h \, {\rm Mpc}^{-1}$, we compute covariance matrices and summary responses to changes in parameters by means of Monte Carlo averaging and numerical differentiation, using the publicly available \Quijote\ and \QuijotePNG\ N-body simulation suites. This a set of simulations with $512^3$ dark matter particles in boxes of size $1~h^{-1}$Gpc.

The paper is structured as follows. In Sect.~\ref{sec:summaries} we introduce the summary statistics considered in our analysis: power spectrum, bispectrum and WST; in Sect.~\ref{sec:fisher} we describe our methodology to compute Fisher matrices; in Sect.~\ref{sec:simulations} we describe in more detail the input \Quijote\ simulation dataset; in Sect.~\ref{sec:results} we describe our analysis and present the main results; in Sect.~\ref{sec:discussion} we discuss the interpretation of our results and draw a comparison between the P$+$B and WST performance; in Sect.~\ref{sec:conclusions} we draw our main conclusions.

\section{Summary Statistics}
\label{sec:summaries}

\subsection{Power spectrum and bispectrum}
\label{sec:p+b}

To extract the cosmological information of the density contrast field $\delta(\vec x)$, we consider its two and three-point correlators in Fourier space, the power spectrum and bispectrum.

From a Fourier-transformed density field $\delta(\vec k)$ defined on a three-dimensional grid, the power spectrum $P(k)$ is easily obtained using a binned estimator (see e.g., \cite{Feldman:1993ky}):
\begin{equation}
    \label{eq:P(k)}
    \hat{P}(k_i) = \frac{1}{V N_i} \sum\limits_{k\in\Delta_i} \delta(\vec k) \delta^*(\vec k),
\end{equation}
where $V$ is the surveyed volume, $\Delta_i$ defines the binning of the relevant $k$-range, $N_i$ is the count of modes $k$ in the $i$-th bin and $k_i$ the average modulus.

Rather than computing the binned bispectrum $B(k_1, k_2, k_3)$, we use a modal approach based on its expansion on a basis of separable functions (i.e.\ symmetrized product of one-dimensional functions in $k$) \citep{Fergusson:2009nv, Fergusson:2010dm, Fergusson:2010ia, Schmittfull:2012hq}. Then, evaluating the information content of the bispectrum only comes to estimating modal modes of the following form:
\begin{equation}
    \label{eq:bispectrum}
    \hat{\beta}_n = \frac{1}{V}\int \diff^3 x\, M_p(\vec x)M_q(\vec x)M_r(\vec x), 
    \quad\text{with}~~M_p(\bx) \equiv \int \frac{\diff^3 k}{(2\pi)^3} \frac{q_p(k)\delta(\vec k)}{\sqrt{k P(k)}}e^{i\vec k \cdot \vec x},
\end{equation}
where the $q_p(k)$ are the chosen one-dimensional functions for the bispectrum expansion. As was shown in \cite{Hung:2019ygc, Byun:2020rgl, Jung:2022rtn}, only a small number of such functions, constructed from simple polynomials (e.g.\ Legendre polynomials) or from the tree-level matter bispectrum, leading to typically only $100$ modal modes $\beta_n$, is necessary to study the matter and halo bispectrum from linear to non-linear scales ($\kmax=0.5\,\hMpc$).

\subsection{Wavelet Scattering Transform}
\label{sec:wst}

The observables in the WST are the {\it scattering coefficients}, which are built  from nonlinear operations on  $\delta(\vec x)$, and are organized in different orders, or {\it layers}. In this paper, we adopt the implementation of the WST encoded in the \textsc{Kymatio}\footnote{\url{https://www.kymat.io}} package \citep{kymatio}. The zeroth order coefficient is just the spatial average of the modulus of $\delta(\vec x)$ raised to some power $q$,
\beq
S_0^q\equiv \langle \left|\delta(\vec x) \right|^q\rangle\,. 
\eeq
To obtain the higher order coefficients a {\it mother wavelet} function is defined, which, for \textsc{Kymatio}, is 
\begin{equation}\label{eq:mother_wav-1}
    \psi_{0\,l}^m(\vec{x},\,\sigma) \equiv
    \frac{C_l}{(2\pi\sigma^2)^{3/2}} \exp\left[-\frac{x^2}{2 \sigma^2}\right] R_l^m \left(\frac{\vec x}{\sigma}\right)\,,
\end{equation}
where $x\equiv|\vec{x}|$, $\hat{x}\equiv\vec{x}/x$, and
the $R_l^m(\vec r)$ are the {\it regular solid harmonics},
\beq
R_l^m(\vec r)\equiv \sqrt{\frac{4\pi}{2 l+1}} \,r^l \,Y_l^m(\hat{r}),
\eeq
with $Y_l^m(\hat{r})$  the  Laplacian Spherical Harmonics.  The coefficients $C_l$ are given in  appendix \ref{kymatioconv}.
From the mother wavelet, a family of rescaled ones, $\psi_{j\,l}^m$, with $j=1,\,2,\,\cdots$, is generated by successive doublings of the  length scale $\sigma$,
\beq
\psi_{j\,l}^m(\vec{x},\,\sigma)\equiv \psi_{0\,l}^m(\vec{x},\,2^j \sigma)=2^{-3 j}\,\psi_{0\,l}^m(2^{-j}\vec{x},\,\sigma)\,.
\eeq
Following \cite{Valogiannis:2021chp}, in this analysis based on simulation data we will set 
\beq
\sigma=0.8\, \frac{L}{N^{1/3}}\,,
\label{eq:sigmaset}
\eeq
where $L$ is the side of the cubic box and $N$ the number of grid points.

In Fourier space, the wavelets read,
\beq
\tilde \psi_{j\,l}^m(\vec{p},\,\sigma)\equiv\int d^3r\, e^{i \vec p\cdot \vec r} \psi_{j\,l}^m(\vec{x},\,\sigma)=
i^l C_l \, \exp\left[-\frac{(2^j \sigma\,p)^2}{2}\right]
\,R_l^m \left(2^j\sigma \vec p\right)\,,
\eeq
with $p\equiv|\vec{p}|$, $\hat{p}\equiv\vec{p}/p$.

First order coefficients are then defined as
\beq
S_1^q(j,l)\equiv \left\langle \left(\sum_{m=-l}^l\left|\delta(\vec x) \ast \psi_{j\,l}^m(\vec{x},\,\sigma) \right|^2\right)^{\frac{q}{2}}\right\rangle\,,
\label{eq:S1def}
\eeq
where the symbol `$\ast$' indicates convolution in configuration space.
Due to the sum over $m$, the coefficients depend, for fixed $q$, on $j$ and $l$ only. In the following, we will also show results for the special case $q=2$, where the interpretation of these coefficients simplifies considerably. Indeed, in this case, the first order coefficients are directly related to the nonlinear power spectrum $P(p)$,
\begin{align}
S_1^2(j,l)&=\int \frac{d^3 p}{(2\pi)^3}\frac{d^3 p'}{(2\pi)^3}e^{-i \vec x\cdot(\vec p+\vec p')} \langle\tilde \delta(\vec p) \tilde\delta(\vec p')\rangle\, \tilde{\psi_{j\,l}^m}(\vec p)\tilde{\psi_{j\,l}^m}(\vec p')\,,\nonumber\\
&=C_l^2 \int \frac{d^3 p}{(2\pi)^3} W_l^2\left[2^j \sigma p\right]\;P(p) \, \nonumber\\
&=C_l^2\,\langle \delta^2(\vec x;\,j,\,l)\rangle\,,
\label{eq:S12}
\end{align}
where we have used the property of spherical harmonics,
\beq
\frac{4\pi}{2l+1}\sum_{m=-l}^l Y_l^m(\hat p)Y_l^m(-\hat p)=1\,,
\eeq
and defined the Fourier space filter, 
\beq
W_l[z]\equiv z^l\, \exp(-z^2/2)\,,
\label{eq:Wfilt}
\eeq
showing  that, for fixed variance of the Gaussian filter (set by $j$), higher values of $l$ probe larger wavenumbers.
The last line of Eq.~\re{eq:S12} is the space averaged variance of the inverse Fourier transform of the filtered  density field $\tilde \delta(\vec p;j,\,l)\equiv W_l\left[2^j \sigma p\right]\tilde \delta(\vec p)$.

To define second order coefficients, we first define a new nonlinear field,
\beq
U(j_1,\,l)(\vec x)\equiv  \left(\sum_{m=-l}^l\left|\delta(\vec x) \ast \psi_{j_1\,l}^m(\vec{x},\,\sigma) \right|^2\right)^{\frac{1}{2}}\,,
\label{eq:Udef}
\eeq 
and then convolve it again with the wavelet at scale $j_2$, using the same operation as \re{eq:S1def},
\beq
S_2^q(j_2,\,j_1,\,l)\equiv 
\left\langle \left(\sum_{m=-l}^l\left|U(j_1,\,l)(\vec x) \ast \psi_{j_2\,l}^m(\vec{x},\,\sigma) \right|^2\right)^{\frac{q}{2}}\right\rangle\,,
\label{eq:S2def}
\eeq
where, again following \cite{Valogiannis:2021chp}, we make the choice of using the same value for $l$ both for the internal ($j_1$) and for the external ($j_2$) convolution. 

The $q=2$ case now gives
\begin{align}
S_2^2(j_2,\,j_1,\,l)&=C_l^2 \int \frac{d^3 p}{(2\pi)^3} W_l\left[2^{j_2} \sigma p\right]^2\;P_U(p; j_1,\,l) \,, \nonumber\\
&=C_l^2\,\langle U^2(\vec x;\,j_2,\,j_1,\,l)\rangle\,,
\label{eq:S22}
\end{align}
where the power spectrum in the integral now refers to the field 
 $U(j_1,\,l)(\vec x)$,
\beq
\langle \tilde U(j_1,\,l)(\vec p)\,\tilde U(j_1,\,l)(\vec p')\rangle\equiv (2\pi)^3\delta_D(\vec p+\vec p')  P_U(p; j_1,\,l)\,,
\eeq
and the filtered field $U(\vec x;\,j_2,\,j_1,\,l)$ is the inverse Fourier transform of $W_l\left[2^{j_2} \sigma p\right] \tilde U(j_1,\,l)(\vec p)$.

The $q=2$ case turns out to be particularly useful in order to compare the performance of the WST with that of the PS, as all the extra information beyond the nonlinear PS is encoded in scattering coefficients beyond first order. The second order coefficients are just the variances of the filtered fields $U(\vec x: j_2,\,j_1,\,l)$, which therefore contain information on the bispectrum and higher order correlators. 

\begin{figure}
    \centering
    \includegraphics[width=\textwidth]{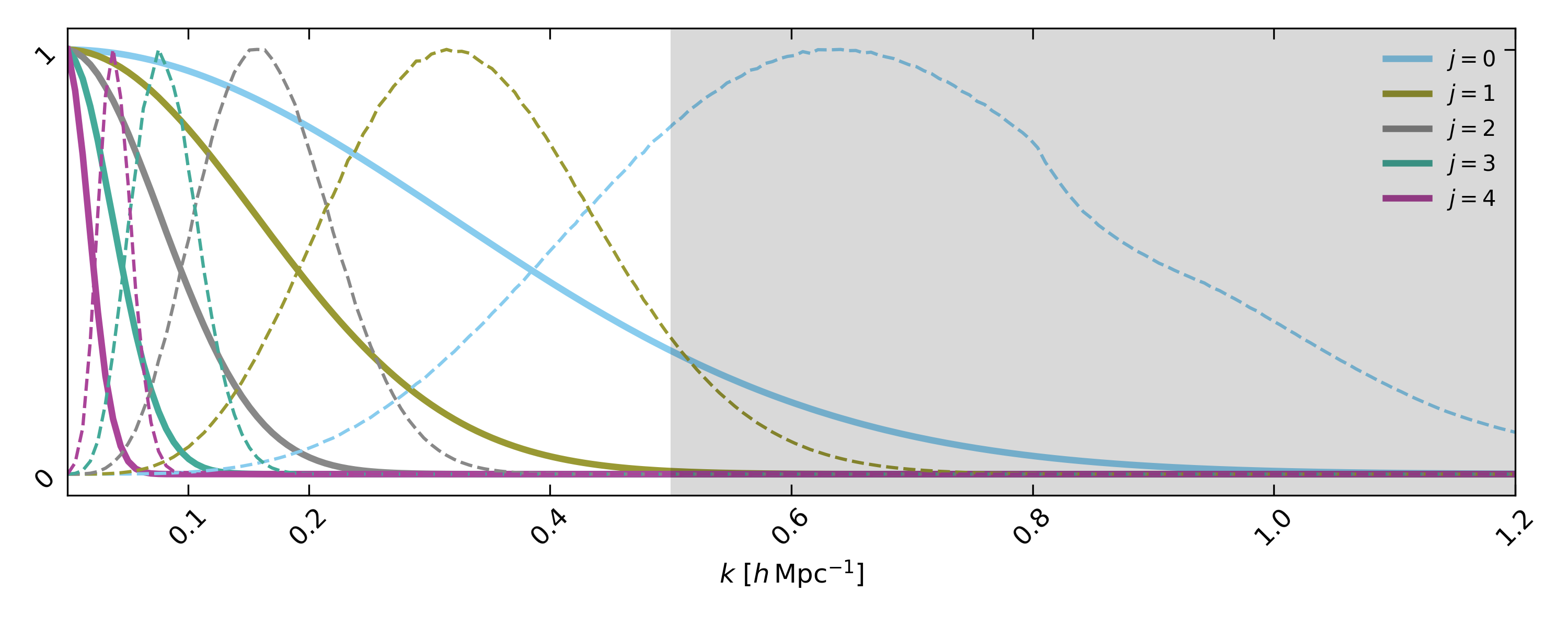}
    \caption{Radial profile of the wavelets at $l=0$ (solid lines) and $l=4$ (dashed lines) at $m=0$ and for different values of $j$, where $j=0$ is equivalent to the mother wavelet. As discussed in the text, we cutoff the field beyond $\kmax = 0.5 \, h \, {\rm Mpc}^{-1}$, this is highlighted by the shaded region.}
    \label{fig:wavelet_coverage-1}
\end{figure}

Fixing $\sigma$ as in Eq.~\eqref{eq:sigmaset}, we must decide how many coefficients to compute at first and second order. For the first order, we take $j=0,\,1,\,\cdots,\,J$, with $J=4$, corresponding to Gaussian smoothing filters in real space of widths from $\sigma$ to $2^4\, \sigma$. In Fig.~\ref{fig:wavelet_coverage-1}, we show the Fourier space filters use to analyze the \Quijote\ suite of simulations presented in Sect.~\ref{sec:simulations}. As for the angular dependence of the wavelets, we take $l=0,\,1\,,\cdots, L$, with $L=4$. When considering second order scattering coefficients, we take the external convolution, controlled by $j_2$, at scales larger than the internal one, therefore $j_1< j_2= 1\,\cdots, J$, with $J=4$ as for the first order coefficients. This gives a total of 76 coefficients, one of zeroth order, 25 of first order and 50 of second order. As we do not consider scattering coefficients at higher orders, these will form our data vector.


\section{Simulation-based Fisher matrices}
\label{sec:fisher}

To benchmark the amount of information on PNG amplitudes, and more generally on cosmological parameters, extracted from the non-linear density field using the different summary statistics presented in Sect.~\ref{sec:summaries}, a standard method is to use the Fisher matrix formalism.

Under the reasonable assumption of normally distributed summary statistics $\bs$, depending on a set of parameters $\btheta$ through their mean $\bar{\bs}$, the Fisher matrix is given by:
\begin{equation}
    \label{eq:fisher}
    F_{ij} = \left(\frac{\partial \bar{\bs}}{\partial \theta_i}\right)^\mathrm{T} \cov^{-1}\left(\frac{\partial \bar{\bs}}{\partial \theta_j}\right),
\end{equation}
where $\cov$ is the covariance matrix. For each parameter $\theta_i$, the quantity $\sqrt{(F^{-1})_{ii}}$ is the Cramer-Rao bound, the expected error bar from an optimal estimator.

To evaluate Eq.~\eqref{eq:fisher}, we choose the simulation-based approach (see Sect.~\ref{sec:simulations} for a description of the simulations used in this work). Unbiased estimates of inverse covariance matrices are obtained using the Hartlap/Anderson correction factor \citep{Hartlap:2006kj}:
\begin{equation}
    \cov^{-1} = \frac{n_\mathrm{r}-n_\mathrm{s}-2}{n_\mathrm{r}-1}\,\hat{\cov}^{-1},\qquad\text{with}~
    \hat{\cov} = \frac{1}{n_r-1} (\bs - \bar{\bs}) (\bs - \bar{\bs})^\mathrm{T},
    \label{eq:covariance}
\end{equation}
where $n_r$ is the number of realizations used and $n_s$ the number of summary statistics. Derivatives can be computed at a chosen fiducial point (denoted by the superscript $*$) by central finite difference:
\begin{equation}
    \label{eq:derivative}
    \left.\frac{\partial \bar{\bs}}{\partial \theta_i}\right|_{\btheta^*} = \frac{\bar{\bs}(\theta_i^* + \delta\theta_i) - \bar{\bs}(\theta_i^* - \delta\theta_i)}{2\delta\theta_i}.
\end{equation}

However, a direct application of this method requires a very large number of simulations to obtain numerically converged results, or can otherwise lead to spuriously super-optimistic Fisher error bars (for a discussion of this issue in the context of PNG, see \cite{Coulton:2022rir, Jung:2022gfa}). This issue can be mitigated using the method of \cite{Coulton:2023sfu}, consisting in computing the Fisher matrix from score-compressed statistics instead of the raw summary statistics. This score-compression is optimal, in the sense it keeps the full information about parameters, under the same conditions as the ones used to write Eq.~\eqref{eq:fisher} for the Fisher matrix \citep{Heavens:1999am, Alsing:2017var}. Score-compressed statistics $\tilde{\bs}$ are obtained using:
\begin{equation}
    \label{eq:compression}
    \tilde{s}_i = \left.\frac{\partial \bar{\bs}}{\partial \theta_i}\right|_{\btheta^*}\cov^{-1} (\bs - \bar{\bs}^*),
\end{equation}
which are the same ingredients as for the Fisher matrix computation, and thus can be evaluated from the same sets of simulations with the advantage of a much faster numerical convergence.


\section{Simulations}
\label{sec:simulations}

Evaluating with precision the statistical properties of the different observables introduced in Sect.~\ref{sec:summaries} up to non-linear scales, as well as their dependence on cosmological parameters, requires a large number of simulations. In this work, we use the \Quijote\footnote{\url{https://quijote-simulations.readthedocs.io/en/latest/}} and \QuijotePNG\footnote{\url{https://quijote-simulations.readthedocs.io/en/latest/png.html}} suites of N-body simulations \citep{Villaescusa-Navarro:2019bje, Coulton:2022qbc} which were designed exactly for this purpose.

These are N-body simulations, containing each $512^3$ dark matter particles in boxes of size $1~{\rm Gpc\,h}^{-1}$, in which the gravitational evolution of structures up to $z=0$ is computed with the TreePM code \textsc{Gadget-III} \citep{Springel:2005mi} starting from initial conditions determined at $z=127$ using \textsc{2LPTIC} \citep{Crocce:2006ve} or \textsc{2LPTPNG} \citep{Scoccimarro:2011pz, Coulton:2022qbc}.\footnote{\url{https://github.com/dsjamieson/2LPTPNG}} The suite contains $15\,000$ simulations with \Planck-like cosmological parameters \citep{Planck:2018vyg}, and subsets of $500$ were one of the parameters is slightly shifted above or below its fiducial value. The characteristics of these different sets are recalled in Table~\ref{tab:quijote}. 

\begingroup
\setlength{\tabcolsep}{4pt} 
\renewcommand{\arraystretch}{1.2} 
\begin{table}
    \begin{center}
    \begin{tabular}{|c|c|cccccccc|c|}
    \hline
& $N_\mathrm{sims}$ & $\sigma_8$ & $\Omega_m$ & $\Omega_{\rm b}$ & $n_s$ & $h$ & $\fNLloc$ & $\fNLeq$ & $\fNLort$ & $M_\mathrm{min} (M_\odot/h)$ \\
\hline
Fiducial & $15000$ & $0.834$ & $0.3175$ & $0.049$ & $0.9624$ & $0.6711$ & $0$ & $0$ & $0$ & $3.2\times10^{13}$\\
 \hline
$\sigma_8^{+}$ & $500$ & $0.849$ & $0.3175$ & $0.049$ & $0.9624$ & $0.6711$ & $0$ & $0$ & $0$ & $3.2\times10^{13}$\\
$\sigma_8^{-}$ & $500$ & $0.819$ & $0.3175$ & $0.049$ & $0.9624$ & $0.6711$ & $0$ & $0$ & $0$ & $3.2\times10^{13}$\\
$\Omega_m^{+}$ & $500$ & $0.834$ & $0.3275$ & $0.049$ & $0.9624$ & $0.6711$ & $0$ & $0$ & $0$ & $3.2\times10^{13}$\\
$\Omega_m^{-}$ & $500$ & $0.834$ & $0.3075$ & $0.049$ & $0.9624$ & $0.6711$ & $0$ & $0$ & $0$ & $3.2\times10^{13}$\\
$n_s^{+}$ & $500$ & $0.834$ & $0.3175$ & $0.049$ & $0.9824$ & $0.6711$ & $0$ & $0$ & $0$ & $3.2\times10^{13}$\\
$n_s^{-}$ & $500$ & $0.834$ & $0.3175$ & $0.049$ & $0.9424$ & $0.6711$ & $0$ & $0$ & $0$ & $3.2\times10^{13}$\\
$h^{+}$ & $500$ & $0.834$ & $0.3175$ & $0.049$ & $0.9624$ & $0.6911$ & $0$ & $0$ & $0$ & $3.2\times10^{13}$\\
$h^{-}$ & $500$ & $0.834$ & $0.3175$ & $0.049$ & $0.9624$ & $0.6511$ & $0$ & $0$ & $0$ & $3.2\times10^{13}$\\
 \hline
$f_\mathrm{NL}^{\mathrm{local},+}$ & $500$ & $0.834$ & $0.3175$ & $0.049$ & $0.9624$ & $0.6711$ & $+100$ & $0$ & $0$  & $3.2\times10^{13}$\\
$f_\mathrm{NL}^{\mathrm{local},-}$ & $500$ & $0.834$ & $0.3175$ & $0.049$ & $0.9624$ & $0.6711$ & $-100$ & $ 0$ & $0$ & $3.2\times10^{13}$ \\
$f_\mathrm{NL}^{\mathrm{equil},+}$ & $500$ & $0.834$ & $0.3175$ & $0.049$ & $0.9624$ & $0.6711$ & $0$ & $+100$ & $0$ & $3.2\times10^{13}$\\
$f_\mathrm{NL}^{\mathrm{equil},-}$ & $500$ & $0.834$ & $0.3175$ & $0.049$ & $0.9624$ & $0.6711$ & $ 0$ & $-100$ & $0$ & $3.2\times10^{13}$ \\
$f_\mathrm{NL}^{\mathrm{ortho},+}$ & $500$ & $0.834$ & $0.3175$ & $0.049$ & $0.9624$ & $0.6711$ & $0$ & $0$ & $+100$ & $3.2\times10^{13}$\\
$f_\mathrm{NL}^{\mathrm{ortho},-}$ & $500$ & $0.834$ & $0.3175$ & $0.049$ & $0.9624$ & $0.6711$ & $ 0$ & $0$ & $-100$ & $3.2\times10^{13}$ \\
 \hline
$M_\mathrm{min}^{+}$ & $500$ & $0.834$ & $0.3175$ & $0.049$ & $0.9624$ & $0.6711$ & $0$ & $0$ & $0$ & $3.3\times10^{13}$\\
$M_\mathrm{min}^{-}$ & $500$ & $0.834$ & $0.3175$ & $0.049$ & $0.9624$ & $0.6711$ & $0$ & $0$ & $0$ & $3.1\times10^{13}$\\
\hline
        \end{tabular}
      \end{center}
\caption{The parameters of the \Quijote~and \QuijotePNG~N-body simulations and halo catalogues used in this work.}
 \label{tab:quijote}
\end{table}

In this work, we focus on the parameters $\vec{\theta}=\{\sigma_8,\,\Omega_m,\,n_s,\,h,\,\fNLloc,\,\fNLeq,\,\fNLort\}$, and study both the matter (at $z=1$) and halo (at $z=0$) density fields. The definitions of the PNG parameters, $\fNLloc,\,\fNLeq,$ and $\fNLort$, characterizing different shapes of the primordial bispectrum, are given in \cite{Coulton:2022qbc}. The halos have been identified using friends-of-friends (FOF) \citep{1985ApJ...292..371D}. Note that we only include halos with a mass above $\Mmin = 3.2\times 10^{13} M_\odot/h$, and include $\Mmin$ as a nuisance parameters by varying it in the halo analyses. As shown in previous analyses \cite{Coulton:2022qbc,Jung:2022gfa}, $\Mmin$ can be thought of as a proxy for the linear bias parameter $b_1$.

We use the power spectra and bispectrum modal modes which were previously measured in \cite{Jung:2022rtn} and \citep{Jung:2022gfa} for the matter and halo fields, respectively. Note that the matter density field is studied in real space, while the halo density field is transformed to redshift-space applying the velocity correction along the $x$ direction.

We estimate the WST coefficients from density fields computed by interpolating the dark matter particle or halo positions on regular three-dimensional grids of size $\Ngrid=256$ using the CIC scheme implemented in the code \textsc{Pylians3}.\footnote{\url{https://github.com/franciscovillaescusa/Pylians3}} We apply a sharp, low-pass filter on the input field $\delta(\vec{x})$ that excludes all modes above $\kmax=0.5\,\hMpc$ (that is, the smallest accessible scale with \Quijote\ simulations), and then compute all 76 scattering coefficients.

\section{Results}
\label{sec:results}

\subsection{First order scattering coefficients and $P(k)$}
\label{sect:1stPS}

As we have shown in Eq.~\re{eq:S12},  first order scattering coefficients with the specific choice $q=2$, that is, $S_1^2(j,\,l)$, contain no more information than the (nonlinear) power spectrum $P(k)$.  To check this explicitly, we run a Fisher matrix analysis on the non-linear dark matter field at redshift $z=1$. The resulting constraints for the set of cosmological parameters $\vec{\theta}=\{\sigma_8,\,\Omega_m,\,n_s,\,h\}$ are shown in Fig.~\ref{fig:WST_vs_Pk_fisher-2}. The near to perfect overlap between the 2D contours for $P(k)$ and $S_1^2(j,\,l)$ shows that the information content of the two statistics is nearly identical. $P(k)$ gives slightly better constraints, which shows that the equivalence between the 25 integrals of $P(k)$ of Eq.~\re{eq:S12} and the 78 $P(k)$ bins is not perfect. This can be probably improved by increasing the maximum $j$ and/or $l$.  However, to limit the computational cost, we decide to stick to our choice since, as we will see, including second order coefficients outperforms the power spectrum for all parameters, including PNG ones,  both for matter and for halos. 

We also show the constraints from the first order coefficients with $q=0.8$. In this case, the equivalence with $P(k)$ is lost, and one expects that some information from higher order correlators should leak into first order scattering coefficients. As expected, the contours for the $S_1^{0.8}(j,\,l)$ coefficients differ from the ones for the $S_1^2(j,\,l)$ and for the $P(k)$. Moreover, for all the parameters but $n_s$, the performance is actually worse, suggesting that the redistribution of the information in passing from $q=2$ to $q=0.8$ moves it from first order to higher order  coefficients.

\begin{figure}
    \centering
    \includegraphics[width=\textwidth]{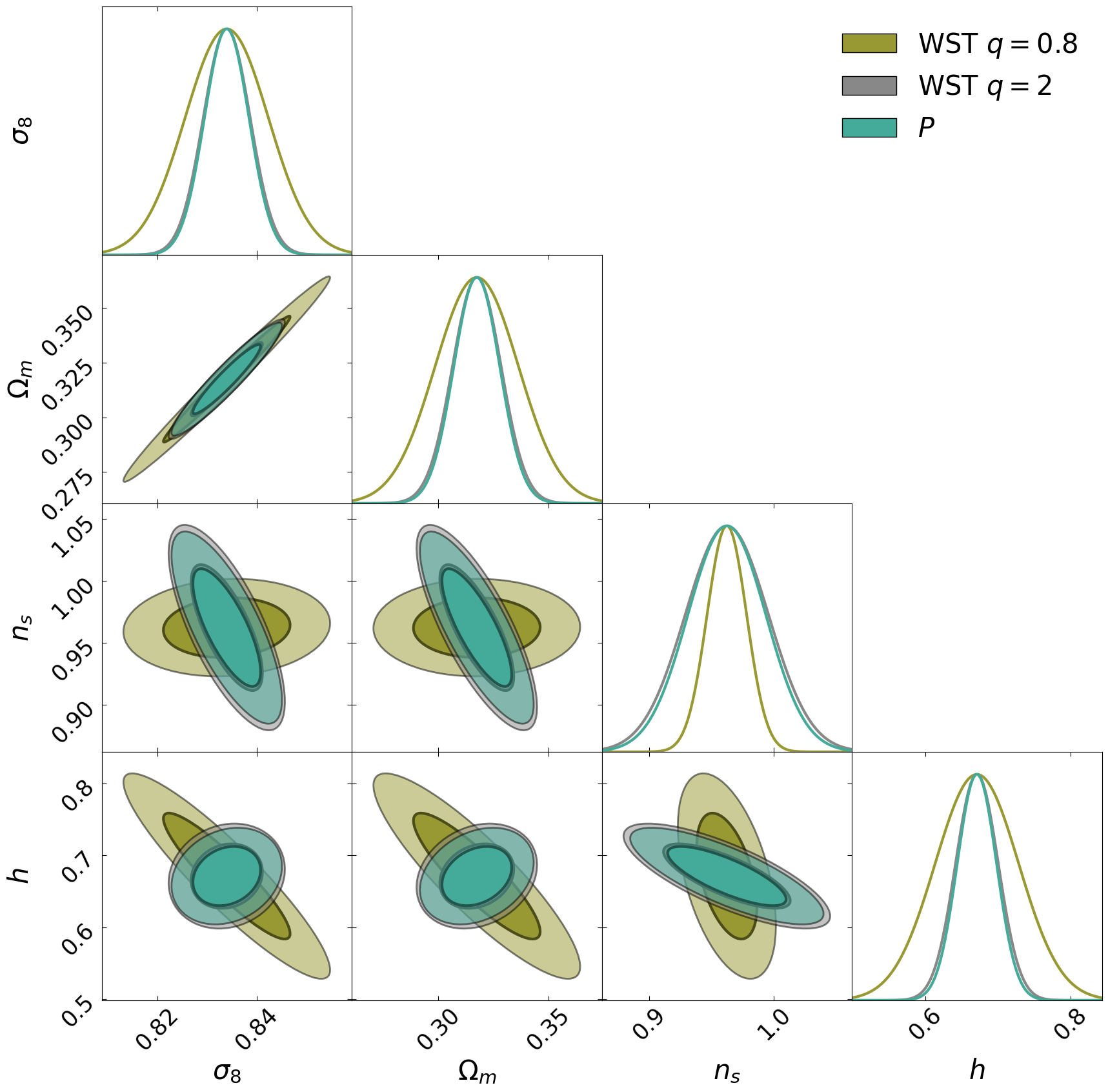}
    \caption{Expected distribution of the parameters around the fiducial \textsc{quijote}-PNG cosmology, with covariance given by the Fisher matrix computed on the non-linear matter field at $z=1$. The statistics compared in the figure are 1st order WST coefficients $S_1^{0.8}(j,\,l)$ (olive), $S_1^2(j,\,l)$ (grey) and $P(k)$ (aquamarine). }
    \label{fig:WST_vs_Pk_fisher-2}
\end{figure}

\subsection{Matter at $z=1$}
\label{sec:matter}
We now consider the matter field at $z=1$ and extend the analysis of the previous subsection by enlarging the parameter vector to include the PNG parameters, and by including second order scattering coefficients and the bispectrum. Specifically, in Figs.~\ref{fig:WST_vs_P+B_fisher_m_q=0.8} and \ref{fig:WST_vs_P+B_fisher_m_q=2} we use the parameter vector
$\vec{\theta}=\{\sigma_8,\,\Omega_m,\,n_s,\,h,\,\fNLloc,\,\fNLeq,\,\fNLort\}$, whereas in the upper table of Fig.~\ref{fig:bounds}, each PNG shape is analyzed jointly with cosmological parameters and independently from the
two others PNG shapes. For cosmological parameters, we report the largest bound of the three
analyses (differences being small anyway).

In Figs.~\ref{fig:WST_vs_P+B_fisher_m_q=0.8} and \ref{fig:WST_vs_P+B_fisher_m_q=2} we show the constraints  from the combination of all the 76 WST coefficients computed with $q=0.8$ and $q=2$, respectively. We compare them with those obtained by  the combination of the power spectrum and the bispectrum ($P+B$), and the combination of all the statistics considered, $P+B+\text{WST}$. The corresponding 1$\sigma$ Cramer-Rao bounds are listed in Fig.~\ref{fig:bounds}. 
We do not show results for $P$ alone, as they are outperformed by both versions of WST on all parameters.  

As discussed in the previous subsection, for $q=2$ the extra information comes entirely from the second order scattering coefficients. For $q=0.8$, the inclusion of second order coefficients improves the performance, which is now overall comparable with that of $q=2$: it is worse for all cosmological parameters except $n_s$, but slightly better for the PNG parameters. 
However, when combined with $P+B$, WST at $q=0.8$ is slightly more efficient than $q=2$ at removing degeneracies between parameters.

\begin{figure}
    \centering
    \includegraphics[width=\textwidth]{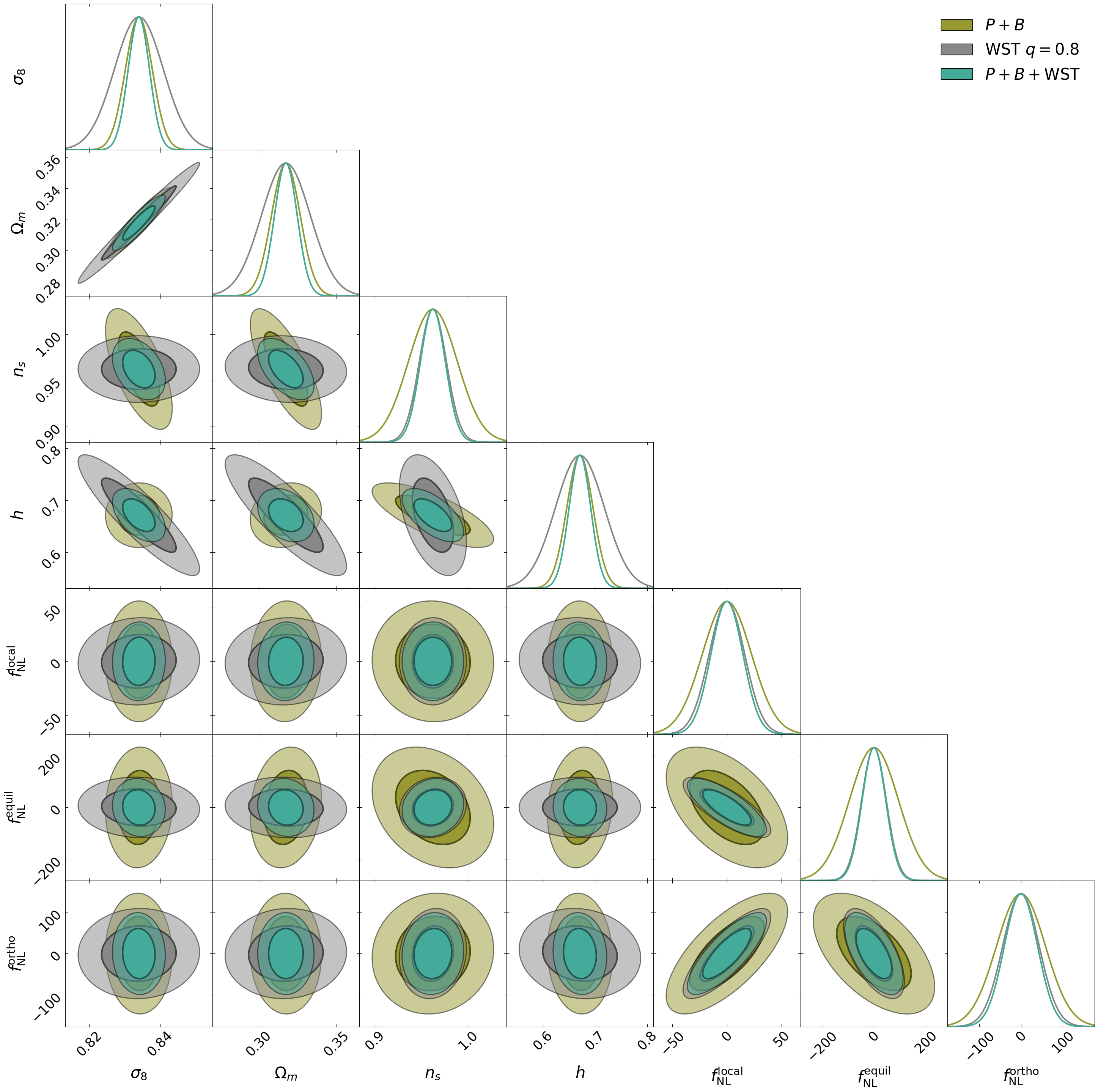}
    \caption{Expected distribution of the parameters around the fiducial \textsc{quijote}-PNG cosmology, with covariance given by the Fisher matrix computed on the non-linear matter field at $z=1$. The statistics compared in the figure are $P+B$ (olive), WST coefficients $S_n^{0.8}(j,\,l)$ with $n=0,\,1,\,2$ (grey), and the full combination $P+B+\mathrm{WST}$ (aquamarine).}
    \label{fig:WST_vs_P+B_fisher_m_q=0.8}
\end{figure}

\begin{figure}
    \centering
    \includegraphics[width=\textwidth]{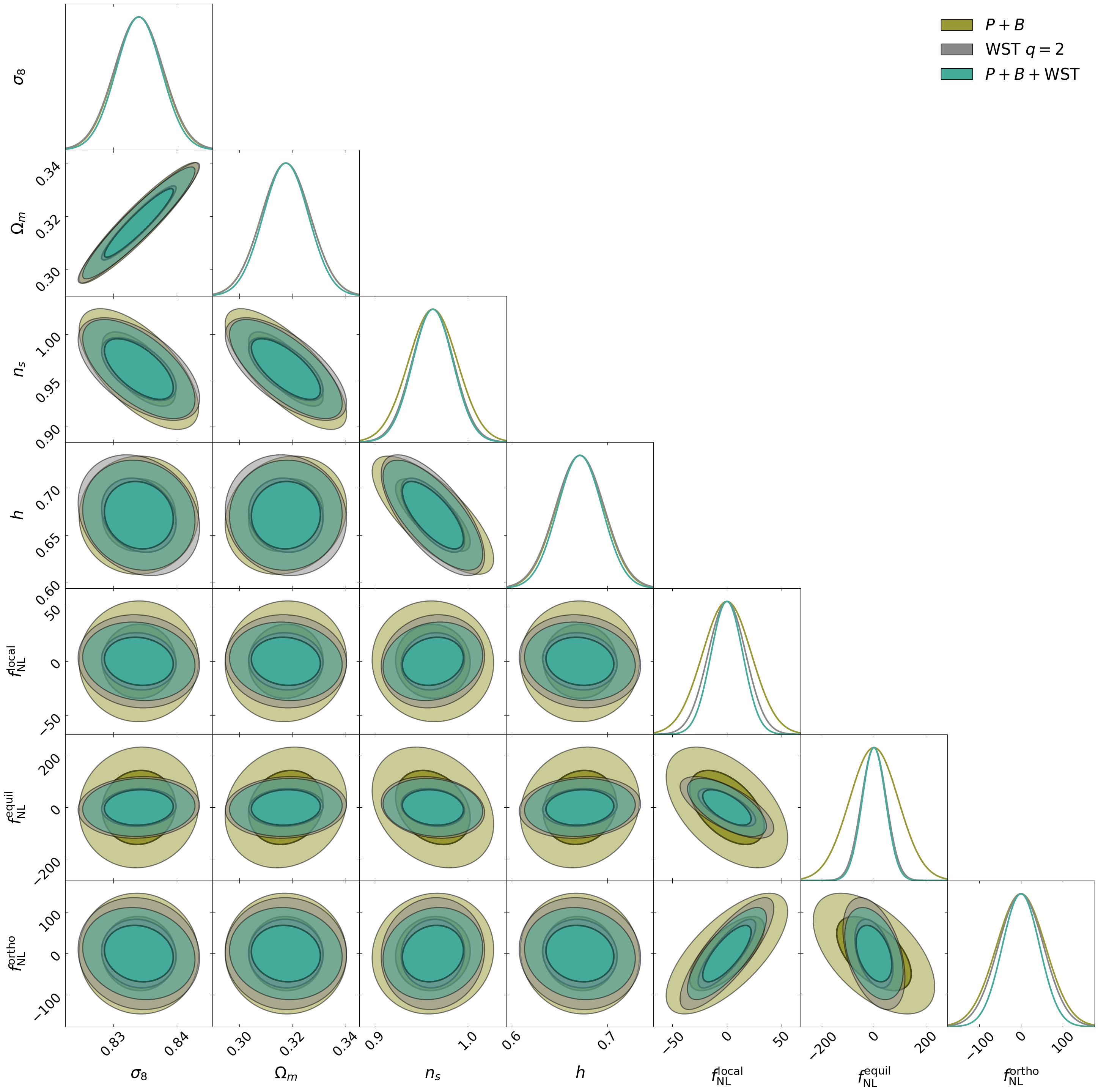}
    \caption{Expected distribution of the parameters around the fiducial \textsc{quijote}-PNG cosmology, with covariance given by the Fisher matrix computed on the non-linear matter field at $z=1$. The statistics compared in the figure are $P+B$ (olive), WST coefficients $S_n^2(j,\,l)$ with $n=0,\,1,\,2$ (grey), and the full combination $P+B+\mathrm{WST}$ (aquamarine).}
    \label{fig:WST_vs_P+B_fisher_m_q=2}
\end{figure}

\subsection{Halos at $z=0$}

We then apply the Fisher matrix analysis on the non-linear dark matter halo field at $z=0$. In order to test the performance of the WST on a biased tracer in a more realistic scenario, the field is transformed in redshift space. The set of cosmological parameters analyzed in Figs.~\ref{fig:WST_vs_P+B_fisher_halo_q=0.8} and \ref{fig:WST_vs_P+B_fisher_halo_q=2} is 
$\vec{\theta}=\{\sigma_8,\,\Omega_m,\,n_s,\,h,\,M_\mathrm{min},\,\fNLloc,\,\fNLeq,\,\fNLort\}$.
We show the bounds given by all the 76 WST coefficients computed with $q=0.8$ and $q=2$ respectively, compared with those given by the first two power spectrum multipoles ($P_0+P_2$), their combination with the bispectrum ($P_0+P_2+B$), and the combination of all the statistics. The  1$\sigma$ Cramer-Rao bounds for all the above statistics are listed in the lower table of Fig.~\ref{fig:bounds}, where each PNG shape is analyzed independently from the other two.

As for matter, the WST outperforms the power spectrum alone for both values of $q$. This is true in particular for the PNG parameters, including for $\fNLloc$ (we do not report the constraints on the other two PNG shapes from $P_0+P_2$, in  Fig.~\ref{fig:bounds}, as they are huge).  $P_0+P_2+B$  performances are comparable with WST for the cosmological parameters (and for $M_{\rm min}$), while for the PNG parameters we find a mixed situation: $\fNLloc$ constraints are better (by about 10-15\%) for $P_0+P_2+B$, $\fNLeq$ are better (by about 20-30\%) for WST, while $\fNLort$ ones are of the same order for WST with $q=2$ and about 30\% better for $q=0.8$.

Overall, our results on halos confirm what we found for matter, namely, that  $q=0.8$ and $q=2$ WST give comparable constraints basically on all parameters. This means  that  taking different exponents does not change significantly the amount of extracted information, once  second order coefficients are included. 

In appendix~\ref{app:marked}, we compare directly the information content on PNG and cosmological parameters of WST with marked statistics, which is very similar.

\begin{figure}
    \centering
    \includegraphics[width=\textwidth]{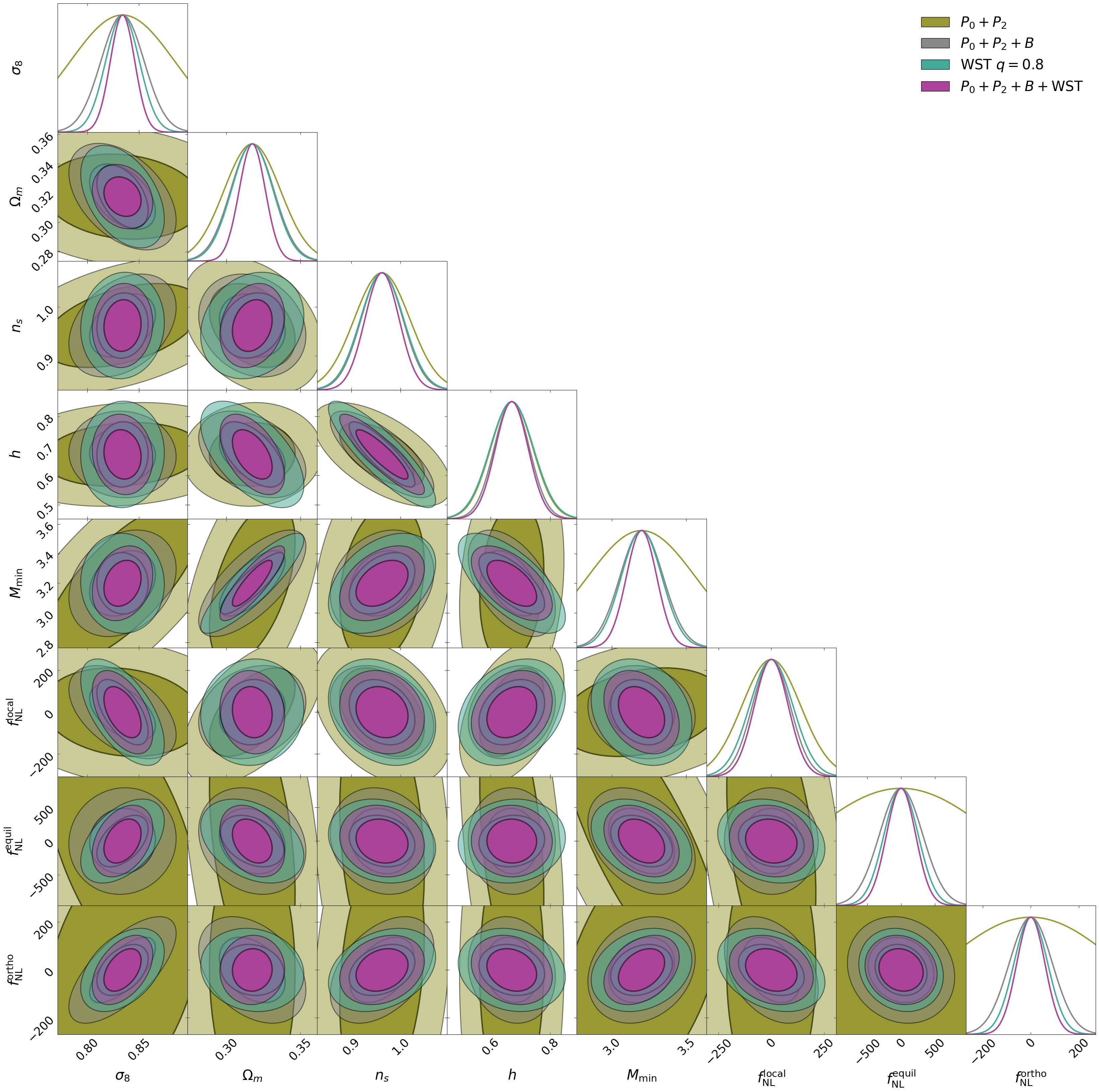}
    \caption{Expected distribution of the parameters around the fiducial \textsc{quijote}-PNG cosmology, with covariance given by the Fisher matrix computed on the non-linear halo field at $z=0$. The statistics compared in the figure are the power spectrum monopole and quadrupole $P_0+P_2$ (olive), $P_0+P_2+B$ (grey), WST coefficients $S_n^{0.8}(j,\,l)$ with $n=0,\,1,\,2$ (aquamarine), and the full combination $P_0+P_2+B+\mathrm{WST}$ (violet).}
    \label{fig:WST_vs_P+B_fisher_halo_q=0.8}
\end{figure}

\begin{figure}
    \centering
    \includegraphics[width=\textwidth]{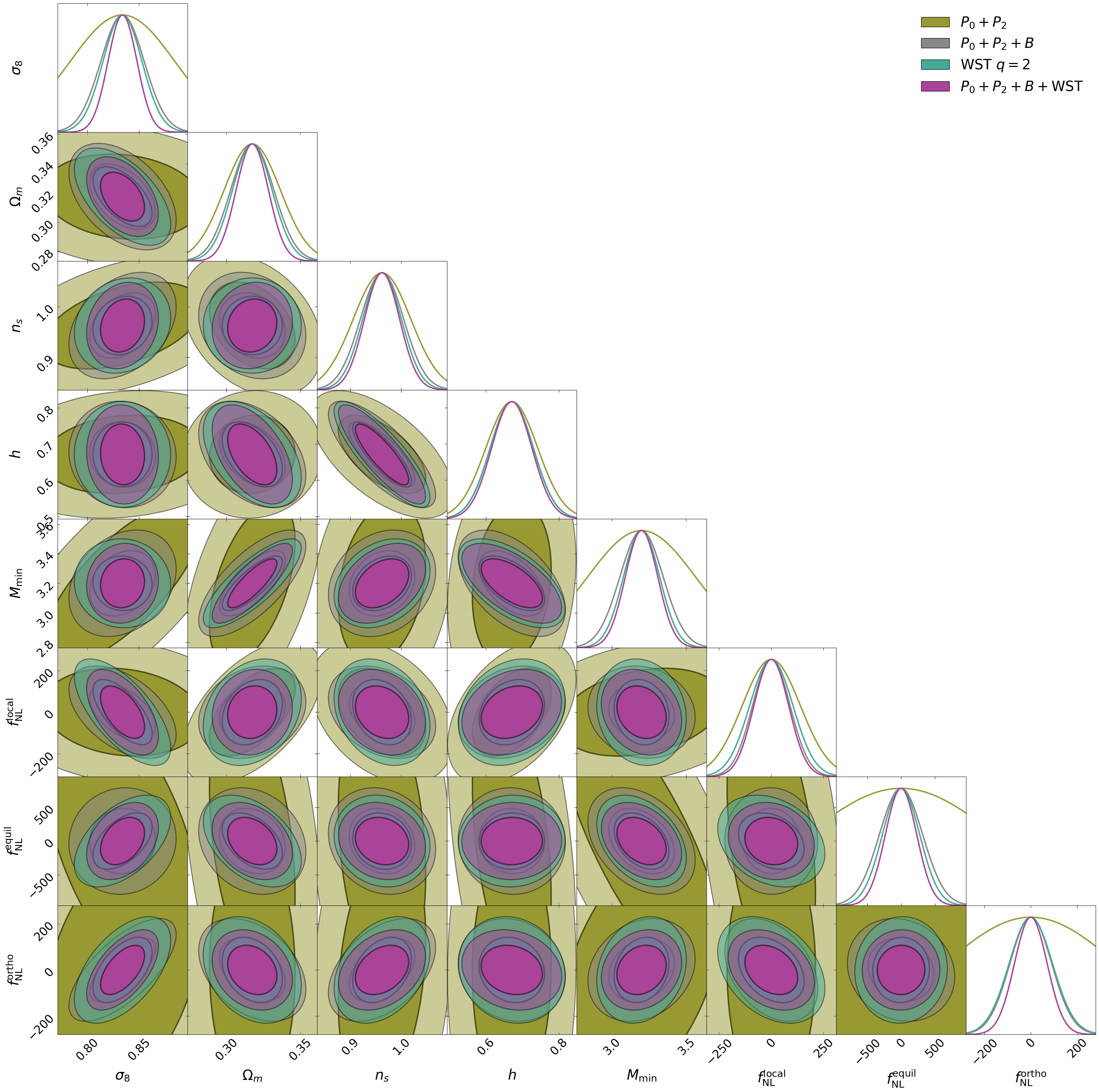}
    \caption{Expected distribution of the parameters around the fiducial \textsc{quijote}-PNG cosmology, with covariance given by the Fisher matrix computed on the non-linear halo field at $z=0$. The statistics compared in the figure are the power spectrum monopole and quadrupole $P_0+P_2$ (olive), $P_0+P_2+B$ (grey), WST coefficients $S_n^2(j,\,l)$ with $n=0,\,1,\,2$ (aquamarine), and the full combination $P_0+P_2+B+\mathrm{WST}$ (violet).}
    \label{fig:WST_vs_P+B_fisher_halo_q=2}
\end{figure}

\begin{figure}
    \centering
    \includegraphics[width=\textwidth]{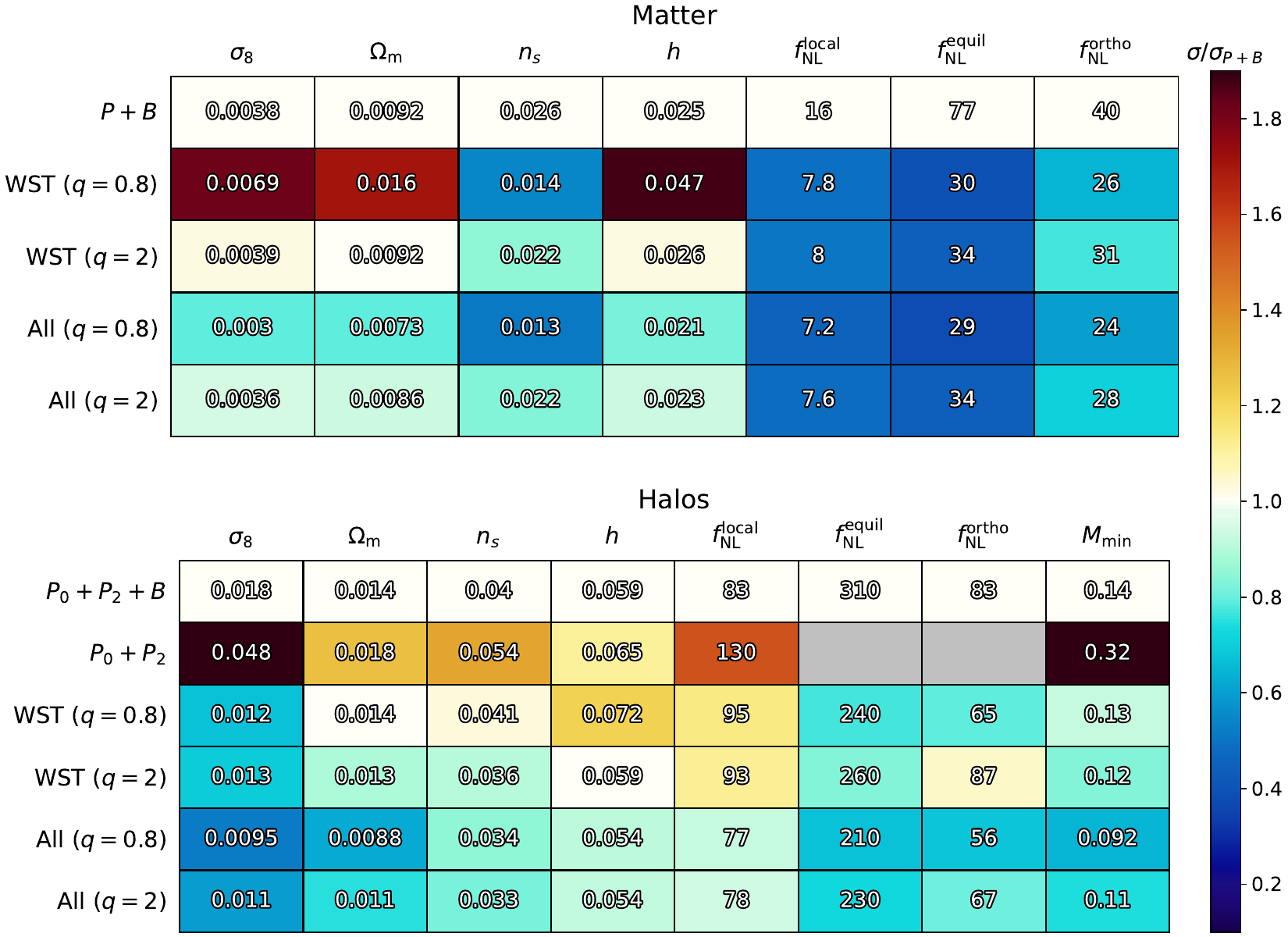}
    \caption{The Cramer-Rao 1$\sigma$ bounds for different combinations of summary statistics (power spectrum, bispectrum and WST, ``all" referring to the three of them altogether). Every statistic is measured up to $\kmax=0.5\,\hMpc$ in the \Quijote\ N-body simulations at $z=1$ (top panel), or in the corresponding halo catalogues at $z=0$ (bottom panel), for a volume of $1\left( h^{-1}{\rm Gpc} \right)^3$. The colour scale indicates the ratio of each error bar with respect to its power spectrum + bispectrum equivalent, highlighting the significant additional information picked by the WST in most cases. Note that each PNG shape is analyzed jointly with cosmological parameters and independently from the two others PNG shapes. For cosmological parameters, we report the largest bound of the three analyses (differences being small anyway).}
    \label{fig:bounds}
\end{figure}

\section{Discussion}
\label{sec:discussion}

\begin{figure}
    \centering
    \includegraphics[width=\textwidth]{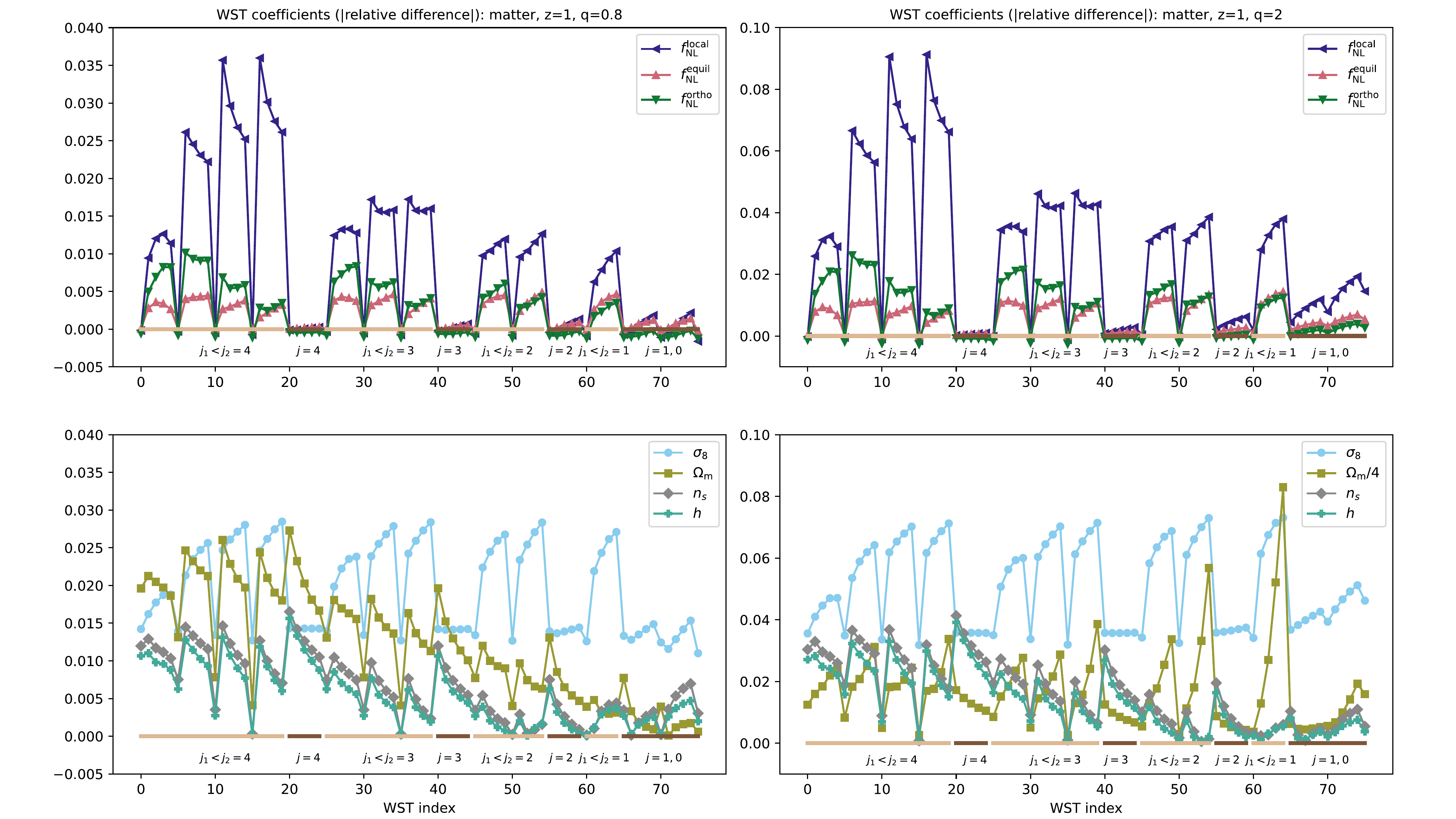}    
    \caption{Variation of WST coefficients for matter in real space with respect to different parameters, normalized to the fiducial values, for the matter field at $z=1$. Left: $q=0.8$, Right: $q=2$. Upper row: NG parameters; lower row: all other parameters. The green segments on the x-axis identify second order WST coefficients, whereas blue segments identify first order ones. There is essentially no NG signal in first order WST coefficients because the power spectrum of the matter field -- to which such coefficients are sensitive -- does not depend on $f_{NL}$ at lowest order. See the main text for details}
    \label{fig:WSTdiff}
\end{figure}

\begin{figure}
    \centering
    \includegraphics[width=\textwidth]{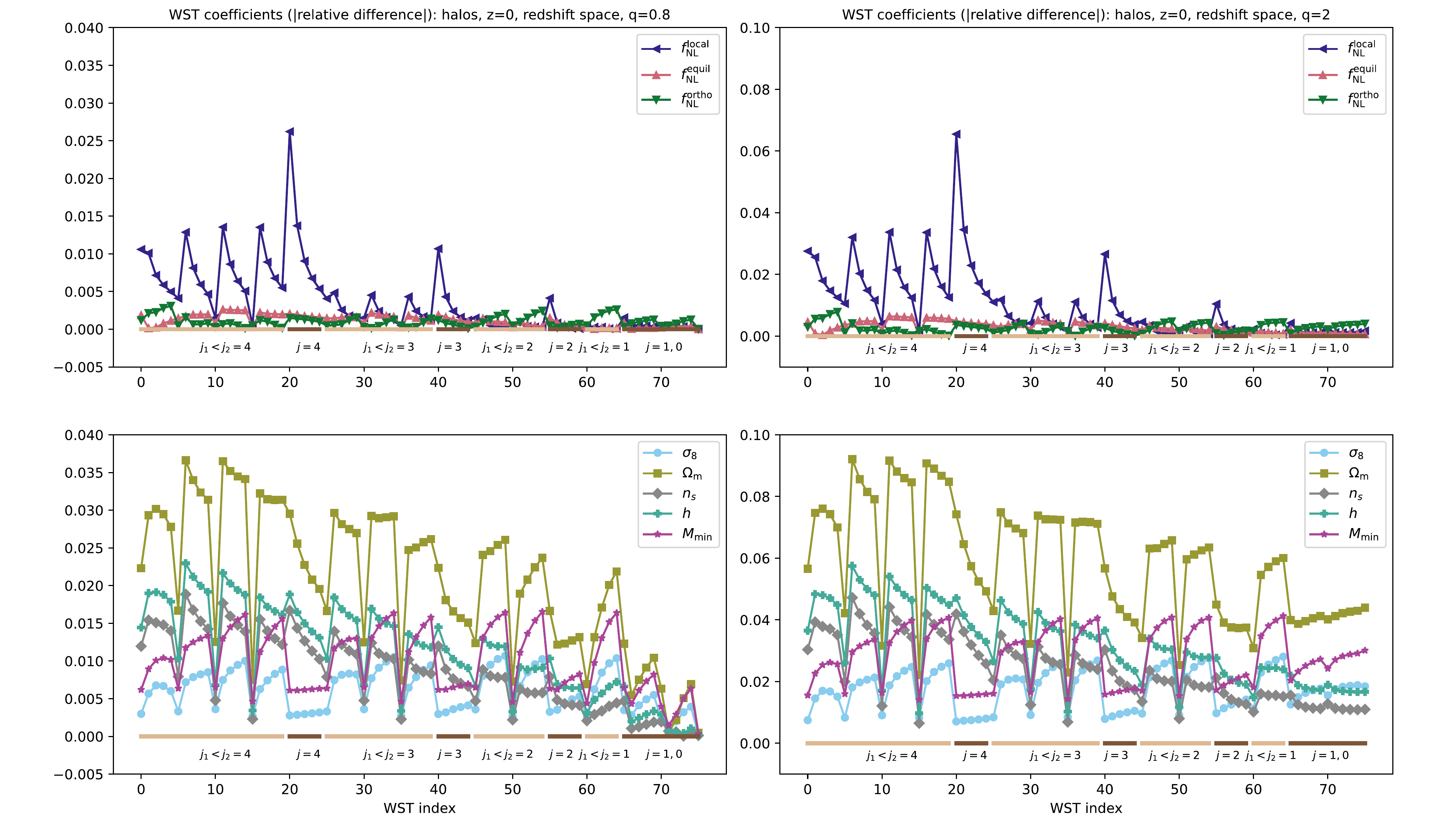}  
    \caption{Variation of WST coefficients for the halo field in redshift space wi th respect to different parameters, normalized to the fiducial values. Now, first order WST coefficients pick up significant primordial NG signal for the local shape, due to the scale dependent bias signature in the power spectrum on large scales. See the main text for more explanations. Left: $q=0.8$, Right: $q=2$.}
    \label{fig:WSTdiffhalos}
\end{figure}
In order to understand where the information on PNG parameters is stored in the WST, we plot in Figs.~\ref{fig:WSTdiff} and ~\ref{fig:WSTdiffhalos} the sensitivity of  WST coefficients to each parameter. More specifically, for each parameter $\alpha$, we plot the absolute value of the difference between the WST coefficients computed in the set of simulations with parameter $\alpha^+$ and the one for $\alpha^-$, normalized to the ones for the fiducial cosmology, (see   Tab.~\ref{tab:quijote} for the specific values taken for each parameter). Fig.~\ref{fig:WSTdiff} is for dark matter at $z=1$ and real space, and Fig.~\ref{fig:WSTdiffhalos} for halos at $z=0$ and redshift space. The upper row of each panels contains PNG parameters, while the lower row all the other ones. Moreover, the left and right columns are obtained by setting the exponent parameter, $q$, to $q=0.8$ and $q=2$, respectively. 

The WST coefficients from 0 to 75 are ordered in decreasing length scale of the most external convolution. So, the first WST coefficient on the left is the second order one with $j_2=4$, $j_1=3$, $l=0$, then  $j_2=4$, $j_1=3$, $l=1$, and so on up to $j_2=4$, $j_1=0$, $l=4$ (end of the first light brown horizontal segment on the right). Then the first order WST coefficients  with $j=4$, $l=0,\cdots 4$ (dark brown segment), followed by second order coefficients with $j_1<j_2=3$, and so on. 

First of all, we notice no clear qualitative difference between left and right columns of each panel, meaning that the intuition developed in sect.~\ref{sec:wst} for the $q=2$ case,  namely, that first order coefficients are related to the power spectrum, extends, to a large extent, also to  $q=0.8$. This explains why the first order coefficients for matter are almost insensitive to PNG parameters (Fig.~\ref{fig:WSTdiff}, upper row, dark brown  segments), especially at large scales. Indeed, the $f_{\rm NL}$ dependence of the matter power spectrum is a suppressed one-loop effect, and therefore it manifests itself only at small scales. The same argument explains  why also second order coefficients with $l=0$ are almost insensitive to PNG parameters, as the sum in $U(j,l)(\vec x)$  of Eq.~\re{eq:Udef} involves just one term and then the second order coefficient is also related to the power spectrum. We conclude that, for matter, the information on PNG parameters comes mainly from second order coefficients with $l\neq 0$. On the other hand, for the other parameters the information is stored also in the power spectrum at linear order, and therefore we expect signal also in first order WST coefficients, as seen in the second row of Fig.~\ref{fig:WSTdiff}.

Before concluding the analysis of the results for matter, we recall that the information content not only depends on the derivatives of the WST coefficients, but also on their covariances. In the first row of  Fig.~\ref{fig:WSTcovariances} we show the correlation coefficients defined as
\beq
r_{ij}\equiv \frac{C_{ij}}{\sqrt{C_{ii} C_{jj}}}\,,
\eeq
where $C_{ij}$ is the covariance between the $i$-th and the $j$-th WST coefficient. $r_{ij}$ can take values between $-1$ (complete anticorrelation) and $1$ (complete correlation), with $r_{ij}=0$ corresponding to no correlation. The  correlation patterns of $q=0.8$ and the $q=2$ cases are clearly different. For the $q=2$ case, the correlation between first order coefficients is very small at large scales and increases at smaller ones, reflecting the behaviour of the power spectrum bins. First order coefficients are also correlated with the second order ones for $l=0$, confirming our interpretation that the latter are also related to the power spectrum. The most striking features are, perhaps, the strong correlation among second order coefficients with $l\neq 0$, and the very small correlation between second order coefficients (with the exception of $l=0$ ones) and first order ones, especially at large scales. This separation is absent for $q=0.8$, confirming our conclusion (see Sect.~\ref{sect:1stPS}) that the PS information is confined to 1st order coefficients for $q=2$ and distributed among 1st and higher order ones for $q\neq 2$. Moreover, the strong correlation between second order coefficients for $q=2$ suggests that a smaller set of them may be enough to extract the same amount of information. 

When we turn to halos, the differences between the PNG sensitivities of first and second order coefficients, and between the correlation patterns for $q=2$ and $q=0.8$, basically disappear for the local shape. In this case, first order WST coefficients are sensitive to $f_{\rm NL}$ since, as is well known, the power spectrum exhibits scale dependent bias at large scales \cite{Dalal:2007cu}. This can be understood as a new contribution to the density field  
\beq
\delta_h=b_1 \delta_m + b_\phi \fNLloc \phi + \epsilon, 
\eeq
where $b_1$ and $ b_\phi$ are tracer-dependent bias parameters (for matter, $b_1=1$ and $b_\phi=0$ ) and $\phi$ is the gravitational potential.
Finally, a stochastic  contribution, $\epsilon$, uncorrelated to $ \delta_m$ and $\phi$, describes `shot noise' in the halo field. It is responsible for the reduction of the sensitivity compared to matter, especially at small scales, and also for the decorrelation between WST coefficients in the $q=2$ case seen in Fig.~\ref{fig:WSTcovariances}.

\section{Conclusions}
\label{sec:conclusions}
In this paper we have investigated the WST as a tool for the study of PNG, and compared its performance with that achievable via a P+B analysis. We analyzed  the matter and  the halo fields, in real and redshift space, measured from the publicly available \Quijote\ and \QuijotePNG\ N-body simulation suite, producing  Fisher forecast for the amplitude parameters, $\fNLloc$, $\fNLeq$, and $\fNLort$, along with the cosmological parameters. 

In order to make a fair comparison, WST and P+B analyses were implemented on the same fields. In particular, to reduce possible spurious effects picked up from WST at small scales, a sharp cut of wavenumbers above $\kmax=0.5\,\;\hMpc$ was applied. 
Moreover, we compared WST's with two values of the exponent parameter $q$, namely, $q=0.8$ and $q=2$, as the latter allows for a more clear interpretation of the results: first order scattering coefficients  are equivalent to the power spectrum, see Fig.~\ref{fig:WST_vs_Pk_fisher-2}, so all information beyond that comes from second order coefficients.

We find that WST's (both for $q=0.8$ and $q=2$) outperform the power spectrum alone on both on PNG and cosmological parameter constraints, as well as on the `bias' parameter $\Mmin$. Specifically, on $\fNLloc$ for halos, the improvement is about 27\%. 
When B is combined with P, halo constraints  from WST are weaker for $\fNLloc$ (at $\sim \, 15\%$ level), but still stronger for $\fNLeq$ ($\sim \, 25\%$)  and $\fNLort$ ($\sim \, 28\%$ for $q=0.8$).

Our results on the standard cosmological (non-PNG) cosmological parameters are in line with previous analyses \cite{Valogiannis:2021chp, Eickenberg:2022qvy}, which also display a better performance of the WST with respect to the power spectrum alone. 
Concerning PNG parameters -- which are the main focus of this work -- we
show that the WST can improve the extraction of PNG information from LSS data over that attainable by a standard P+B analysis. The level of improvement for \QuijotePNG\ simulations is competitive with the best results we obtained in previous works, using other summary statistics, such as the marked power spectrum and bispectrum \cite{2024arXiv240300490J}.

Further developments should advance the current analysis in two directions. On one hand, the level of realism of data should be increased, in order to match as closely as possible that of a realistic survey. Therefore, modeling of galaxy bias, survey systematics, and so on, should be included in our pipeline. On the other hand, advances in the theoretical interpretation of the WST are highly desirable. We have provided a first attempt  in this paper, by considering the WST for $q=2$. This choice enables a cleaner theoretical description, while carrying at the same time basically no information loss with respect to other $q$ values. It will be interesting to investigate more general families of wavelets, in search for those best   tailored for extracting information, especially on PNG parameters, from the LSS. For instance, the angular structure could be exploited to minimize the effect of redshift space distortions, and the underlying symmetries of the problem, such as  extended galileian invariance \cite{Peloso:2013zw,Kehagias:2013yd,Creminelli:2013mca,Peloso:2013spa,Peloso:2016qdr,DAmico:2021rdb}, could be exploited to reduce the number of independent coefficients to take into account in the analyses.

\section*{Acknowledgements}
\noindent
We thank William Coulton, Francisco Villaescusa-Navarro and Ben Wandelt for useful discussions and feedback on the manuscript.
M. Liguori and M. Pietroni  acknowledge support by the MIUR Progetti di Ricerca di Rilevante Interesse Nazionale (PRIN) Bando 2022 - grant 20228RMX4A.
GJ acknowledges support from the ANR LOCALIZATION project, grant ANR-21-CE31-0019 / 490702358 of the French Agence Nationale de la Recherche. 

\begin{figure}
    \centering
    \includegraphics[width=\textwidth]{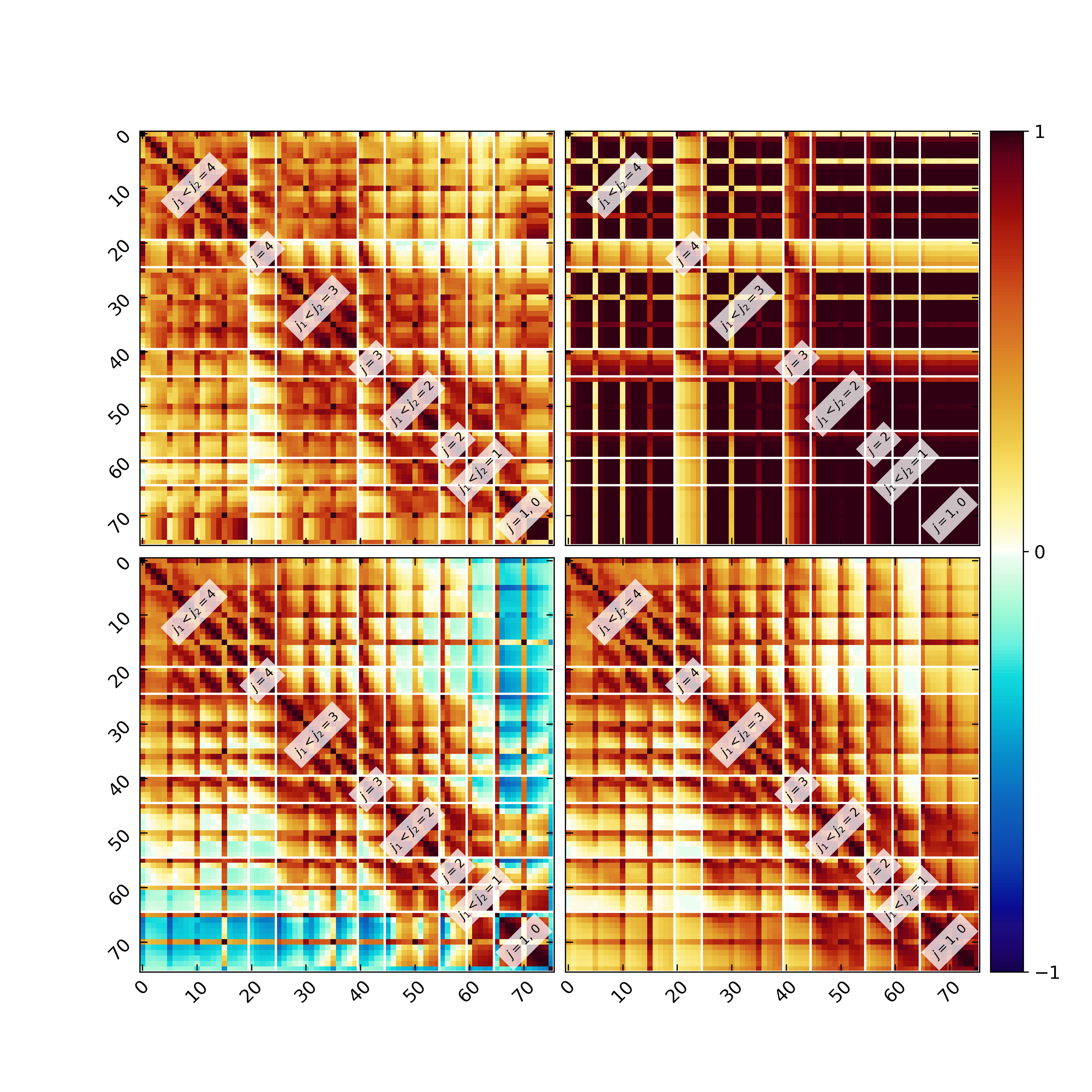}
    \caption{The correlation matrix of the WST coefficients. Upper row: matter, lower row: halos.  Left: $q=0.8$, Right: $q=2$. 
    }
    \label{fig:WSTcovariances}
\end{figure}


\appendix

\section{Comparison with marked statistics}
\label{app:marked}

In \cite{2024arXiv240300490J}, the marked power spectrum and marked bispectrum, computed from re-weighted density fields, were identified as promising observables of PNG on non-linear scales. In Fig.~\ref{fig:spider}, we compare their reported 1$\sigma$ Fisher bounds combining standard and marked power spectra and bispectra to our strongest constraints based on WST reported in Fig.~\ref{fig:bounds}. We obtain very similar constraints on PNG by including marked statistics or the WST, indicating that they probe the same information from higher order correlators beyond the bispectrum.

\begin{figure}
    \centering
    \includegraphics[width=0.99\textwidth]{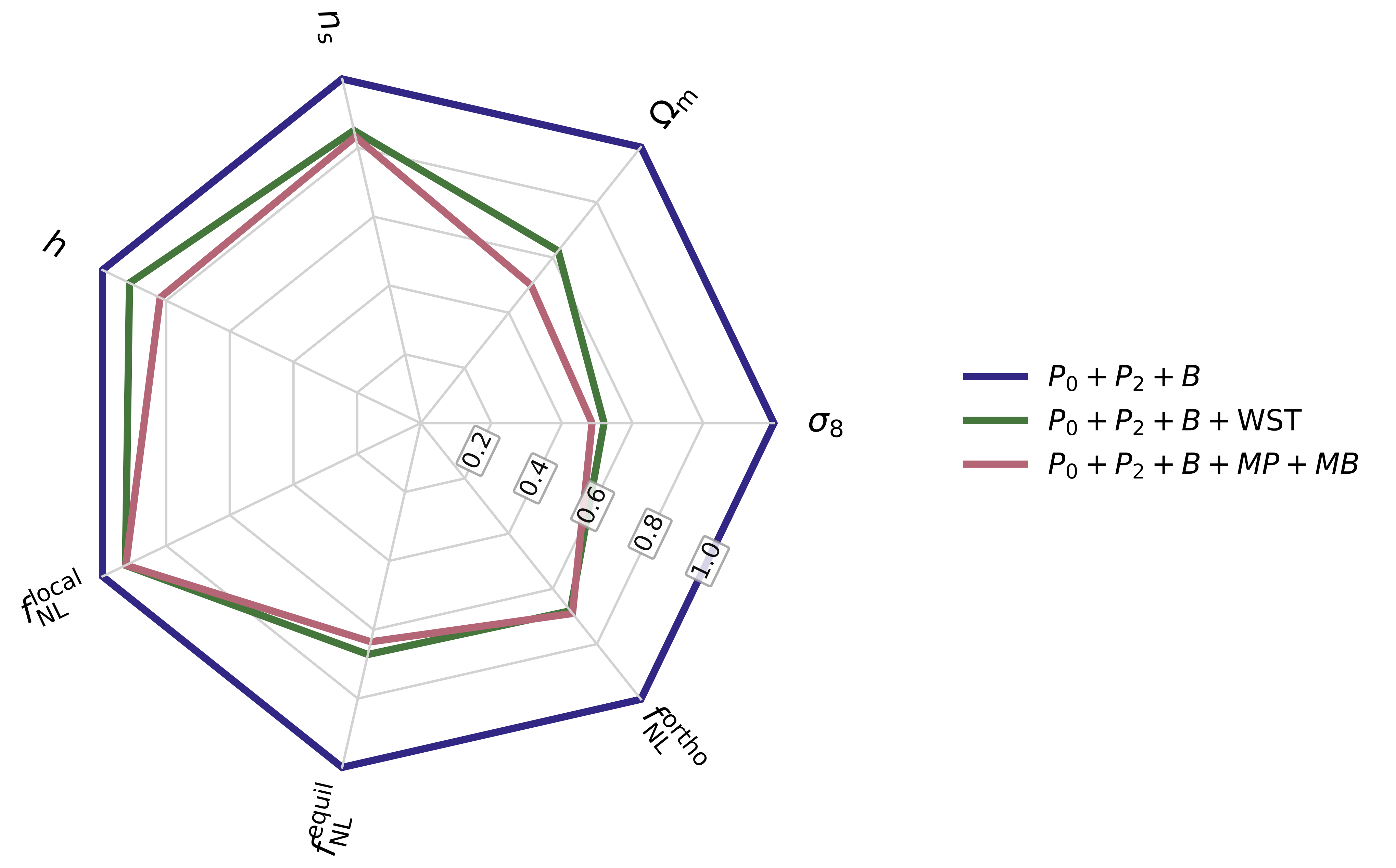}
    \caption{Comparison of the $1$-$\sigma$ Fisher bounds on cosmological parameters and PNG amplitudes for different combinations of summary statistics (power spectrum, bispectrum, WST ($q=0.8$), marked power spectrum and marked bispectrum), measured up to $\kmax=0.5\,\hMpc$ in the \Quijote\ halo catalogues at $z=0$.}
    \label{fig:spider}
\end{figure}

\section{Mother wavelets in \textsc{Kymatio}}
\label{kymatioconv}

The $C_l$ coefficients appearing in \re{eq:mother_wav-1}
are given by 
\beq
C_l=\left\{\begin{array}{cl}
\ds\frac{1}{(l+1)!!}\,,& l\;{\rm even}\\
& \\
\ds\frac{(2 \pi)^{1/2}}{2^{(l+3)/2}\,\left(\frac{l+1}{2}\right)!}\,,& l\;{\rm odd}\,.
\end{array} \right.
\eeq

\section{Convergence and Saturation}

Having obtained our results from a simulation-based approach, we must check that our statistics have converged numerically. We test this by comparing the variation in the Cramer-Rao bounds obtained by using a growing number $N_\mathrm{der}<500$ of simulations to compute the numerical derivative of Eq.~\eqref{eq:derivative}, with the same result given by the full simulation suite (that is, with $N_\mathrm{der}=500$). In figure \ref{fig:WST_convergence} we report the ratio between the two bounds as a function of $N_\mathrm{der}$ for the WST, highlighting the $\fNL$ configurations for clarity. If a line asymptotically converges to $1$, then the bound for that parameter is considered to have numerically converged. We see that the bounds appear to converge rather quickly, especially for $\fNL$ in the case of halos. In the case of matter, the situation is reversed, however the absolute difference in the values never exceeds $3\%$, which implies very high numerical stability. For similar tests, showing the stability of the P + B analysis, see \cite{Coulton:2022qbc, Coulton:2022rir, Jung:2022rtn, Jung:2022gfa}.
It is also interesting to see how  the Cramer-Rao bounds improve  when adding more and more WST coefficients.  This is shown in Fig.~\ref{fig:saturation-9}, where, as in Figs.~\ref{fig:WSTdiff}, \ref{fig:WSTdiffhalos}, and \ref{fig:WSTcovariances}, we order the coefficients starting from $j=4$, all the way to $j=0$ (see Sect.~\ref{sec:discussion}). The   horizontal line at 1 indicates the corresponding bounds from $P+B$. Focusing on PNG parameters, we see that  extra information with respect to $P+B$ comes in already at intermediate scales (starting from the $40^{\rm th}$ coefficient, approximately) and, in the case of halos it shows clear saturation at small scales, supporting the robustness of the bounds with respect to spurious small scale effects.  

\begin{figure}
    \centering
    \includegraphics[width=\textwidth]{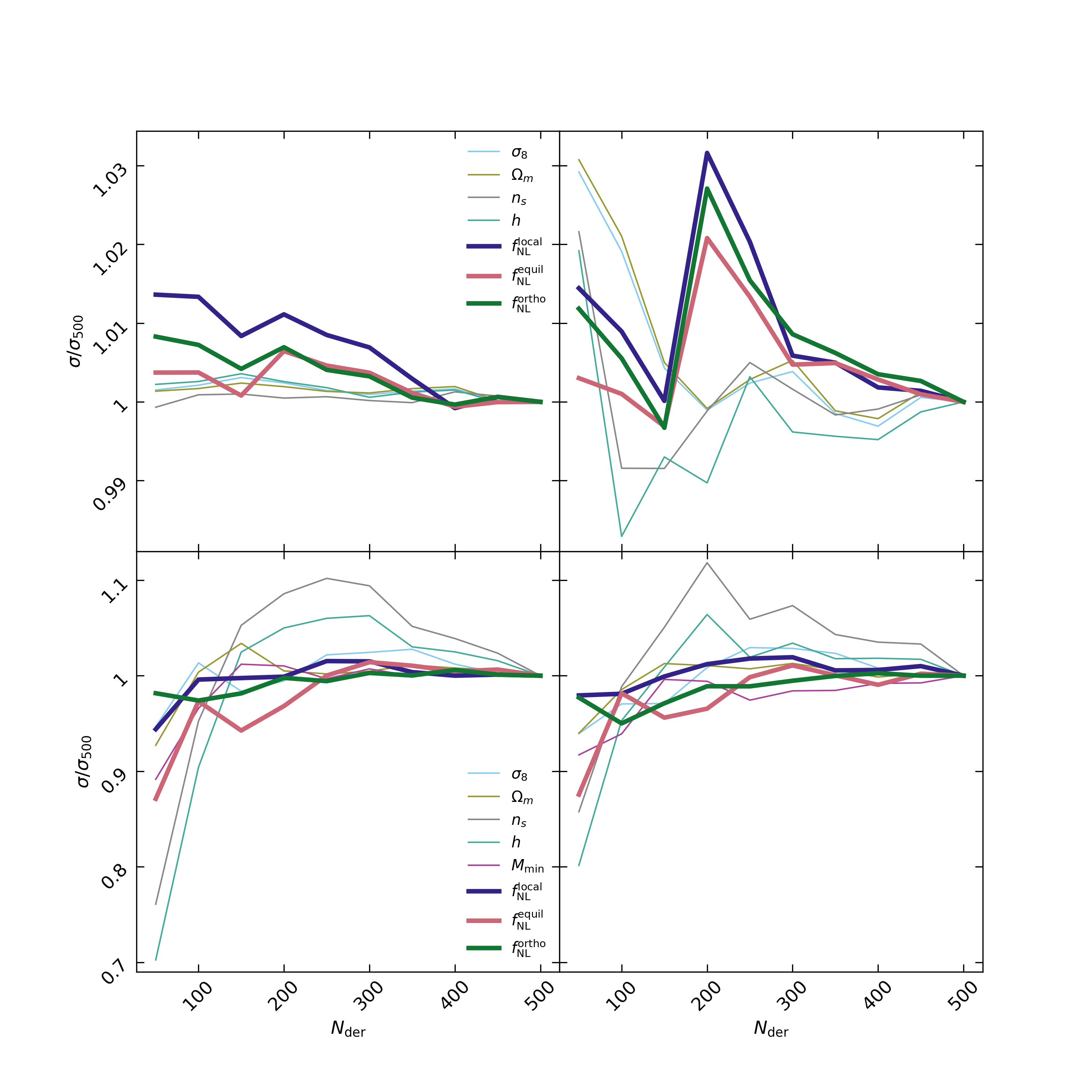}
    \caption{Convergence of the Cramer-Rao 1-$\sigma$ bounds on the cosmological parameters as more derivatives on the WST coefficients are used to compute the Fisher matrix. Upper row: matter, lower row: halos.  Left: $q=0.8$, Right: $q=2$.}
    \label{fig:WST_convergence}
\end{figure}

\begin{figure}
    \centering
    \includegraphics[width=\textwidth]{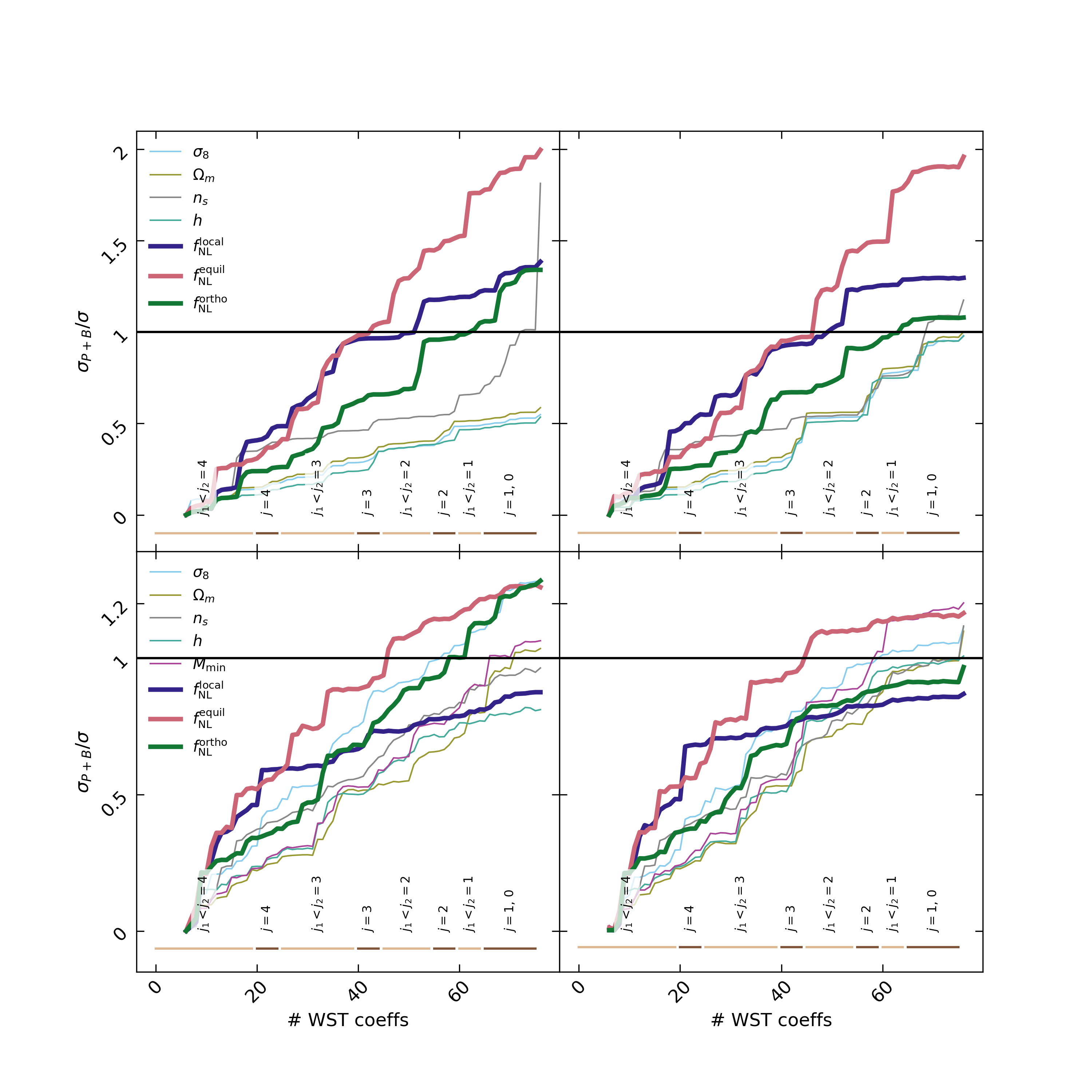}
    \caption{Change in the Cramer-Rao 1$\sigma$ bounds on the cosmological parameters obtained from as we add more WST coefficients to the Fisher matrix. The horizontal lines represent the bounds given by $P+B$. Upper row: matter, lower row: halos.  Left: $q=0.8$, Right: $q=2$.}
    \label{fig:saturation-9}
\end{figure}

\bibliographystyle{JHEP2015}
\bibliography{mybib}
\label{lastpage}

\end{document}